\documentclass[fleqn,twoside]{article}%
\usepackage{amsmath}
\usepackage{amsfonts}
\usepackage{amssymb}
\usepackage{graphicx}%
\usepackage{caption}
\def\be{\begin{equation}}

\def\ee{\end{equation}}
\def\bes{\begin{equation}\begin{split}&}
\def\es{\end{split}}
\def\bi{\bibitem}

\begin{document}
\title{Inflation with scalar-tensor theory of gravity}
\author{Dalia Saha \footnote{E-mail:daliasahamandal1983@gmail.com},~Susmita sanyal\footnote{E-mail:susmitasanyal@yahoo.com}~and Abhik Kumar Sanyal\footnote{E-mail:sanyal\_ak@yahoo.com}}

\maketitle
\noindent
\begin{center}
\noindent
Dept. of Physics, Jangipur College, Murshidabad, West Bengal, India - 742213\\
\end{center}

\noindent

\begin{abstract}
The latest released data from Planck in 2018, put up tighter constraints on inflationary parameters. In the present article, the in-built symmetry of the non-minimally coupled scalar-tensor theory of gravity is used to fix the coupling parameter, the functional Brans-Dicke parameter, and the potential of the theory. It is found that all the three different power-law potentials and one exponential, pass these constraints comfortably, and also gracefully exit from inflation.
\end{abstract}

\noindent
\section{\bf{Introduction}}

The standard (FLRW) model of cosmology based on the basic assumption of homogeneity and isotropy, known as the `cosmological principle', has successfully been able to explain several very important issues in connection with the evolution of the universe. First of all, it predicts the observed expansion of the universe being supported by the Hubble's law. It also postulates the existence of cosmic microwave background radiation (CMBR), formed since recombination when the electrons combined to form atoms, allowing photons to free stream, with extreme precession, being verified by Penzias and Wilson for the first time. It also predicts with absolute precession the abundance of the light atomic nuclei ($\mathrm{{^4He\over H} \sim 0.25, {^2D\over H} \sim 10^{-3}, {^3He\over H} \sim 10^{-4}, {^7Li\over H} \sim 10^{-9}}$, by mass and not by number) observed in the present universe. Finally, assuming the presence of the seeds of perturbation in the early universe, it can explain the observed present structure of the universe. Despite such tremendous success, the model inevitably suffers from a plethora of pathologies. The problems at a glance are the following.\\ 1. `The singularity problem': Extrapolating the FLRW solutions back in time one encounters an unavoidable singularity, since all the physical parameters viz. the energy density ($\rho$), the thermodynamic pressure ($p$), the Ricci scalar ($R$), the Kretschmann scalar ($R_{\alpha\beta\gamma\delta}R^{\alpha\beta\gamma\delta}$) etc. diverge.\\ 2a. `The flatness problem': The model does not provide any explanation to the observed value of the density parameter $\Omega \approx 1$, which depicts that the universe is spatially flat.\\ 2b. `The horizon problem': It also can not provide any reason to the observed tremendous isotropy of the CMBR being split in $1.4\times 10^4$ patches of the sky, that were never causally connected before emission of the CMBR.\\ 2c. `The structure formation problem': It does not also provide any clue to the seeds of perturbation responsible for the structure formation.\\ 3. `The dark energy problem': Finally, the standard FLRW model does not fit the redshift versus luminosity-distance curve plotted in view of the observed SN1a (Supernova type a) data.\\

In connection with the first problem, viz. the so called `Big-Bang singularity' and also to understand the underlying physics of `Black-Hole' being associated with Schwarzschild singularity, it has been realized long ago that 'General Theory of Relativity' (GTR) must have to be replaced by a quantum theory of gravity when and where gravity is strong enough. However, GTR is not renormalizable and a renormalized theory requires to include higher-order curvature invariant terms in the gravitational action. Despite serious and intense research over several decades and formulation of new high energy physical theories like superstring and supergravity theories, a viable quantum theory of gravity is still far from being realized. In connection with last problem, a host of research is in progress over last two decades. It has been realized that to fit the observed redshift versus luminosity-distance curve, it is  either required to take into account some form of exotic matter in addition to the barotropic fluid (ordinary plus the cold dark matter) which violates the strong energy condition ($\rho + p \ge 0,~\rho + 3p \ge 0$), and is dubbed as `dark energy' (since it interacts none other than with the gravitational field) or to modify the theory of gravity by including additional curvature scalars in the Einstein-Hilbert action, known as `the modified theory of gravity'. It has been observed that both the possibilities lead to present accelerated expansion of the universe. The problem is thus rephrased as: why the universe undergoes an accelerated expansion at present? The pathology 2, in connection with the flatness, horizon and structure formation problems has however been solved under the hypothesis called `Inflation', which is our present concern.\\

\noindent
Under the purview of cosmological principle, i.e. taking into account Robertson-Walker line element,

\be \label{1.1} ds^2 = - dt^2 + a^2(t) \left[\frac{dr^2}{1-kr^2} + r^2 (d\theta^2 + sin^2 \theta d\phi^2)\right],\ee
the co-moving distance (the present-day proper distance) traversed by light between cosmic time $t_1$ and $t_2$ in an expanding universe may be expressed as, $d_0(t_i, t_f) = a_0\int_{t_i}^{t_f} {dt\over a(t)}$, where $a(t)$ is called the scale factor. The co-moving size of the particle horizon at the last-scattering surface of CMBR ($a_f  = a_{lss}$) corresponds to $d_0 \sim 100$ Mpc, or approximately $1^0$ (one degree) on the CMB sky today. In the decelerated radiation dominated era of the standard model of cosmology (FLRW model), for which $a \propto \sqrt t$ the integrand, $(\dot a)^{-1} \sim 2\sqrt t$ decreases towards the past, and there exists a finite co-moving distance traversed by light since the Big Bang ($a_i \rightarrow 0$), called the particle horizon. The hypothesis of inflation \cite{1,2} postulates a period of accelerated expansion, $\ddot a > 0$, in the very early universe, prior to the radiation-dominated era, administering certain initial conditions \cite{3,4,5,6,7,8,9}. During a period of inflation e.g. a de-Sitter universe ($a \propto e^{\Lambda t}$) driven by a cosmological constant (say), $(\dot a)^{-1}\sim (\Lambda e^{\Lambda t} )^{-1}$  increases towards the past, and hence the integral diverges as ($a_i \rightarrow 0$). This allows an arbitrarily large causal horizon dependent only upon the duration of the accelerated expansion. Assuming that the universe inflates with a finite Hubble rate $H_i$, (instead of a constant exponent $\Lambda$) ending with $H_f  < H_i$, we may have, $d_0 (t_i,t_f )>({a_i\over a_f}) H_i^{-1} (e^{N}-1)$ where $N = \ln\Big({a_{f}\over a_i }\Big)$ is measured in terms of the logarithmic expansion (or `e-folds'), and describes the duration of inflation. It has been found that a $40-60$ e-folds of inflation can encompass our entire observable universe today, and thus solves the horizon and the flatness problem discussed earlier. In some situations, e-fold may range between $25 \le N \le 70$, depending on the model under consideration.\\

A false vacuum state can drive an exponential expansion, corresponding to a de-Sitter space-time with a constant Hubble rate on spatially-flat hypersurfaces. However, a graceful exit from such exponential expansion requires a phase transition to the true vacuum state. A second-order phase transition \cite{10,11}, under the slow roll condition of the scalar field (that can also drive the inflation instead of the cosmological constant), potentially leads to a smooth classical exit from the vacuum-dominated phase. Further, the quantum fluctuations of the scalar field, which essentially are the origin of the structures seen in the universe today, provides a source of almost scale-invariant density fluctuations \cite{12,13,14,15,16}, as detected in the CMBR. Accelerated expansion and primordial perturbations can also be produced in some modified theories of gravity (e.g., \cite{1,17} and also a host of models presently available in the literature), which introduce additional non-minimally coupled degrees of freedom. Such inflationary models are conveniently studied by transforming variables to the so-called ‘Einstein frame’, in which Einstein’s equations apply with minimally coupled scalar fields \cite{18,19}, which we shall deal with, in the present manuscript.\\

Non-minimal coupling with the scalar field $\phi$ is unavoidable in a quantum theory, since such coupling is generated by quantum corrections, even if it is primarily absent in the classical action. Particularly, it is required by the renormalization properties of the theory in curved space-time background. Recently, in view of a general conserved current, obtained under suitable manipulation of the field equations \cite{cons1, cons2, cons3, cons4}, a non-minimally coupled scalar-tensor theory of gravity has been studied extensively in connection with the cosmological evolution, starting from the very early stage (Inflationary regime) to the late-stage (presently accelerated matter-dominated era) via a radiation dominated era \cite{Beh}. It has been found that such a theory admits a viable inflationary regime, since the inflationary parameters viz. the scalar-tensor ratio ($r$), and the spectral index $n_s$ lie well within the limits of the constraints imposed by Planck's data, released in 2014 \cite{Planck14} and in 2016 \cite{Planck16}. Further the model passes through a Friedmann-like radiation era ($a\propto \sqrt t, q = 1$,) and also an early stage of long Friedmann-like decelerating matter dominated era ($a\propto t^{2\over 3},~ q = {1\over 2}$) till $z \approx 0.4$, where $a$, $q$ and $z$ denote the scale factor, the deceleration parameter and the red-shift respectively. The universe was also found to enter a recent accelerated expansion at a red-shift, $z \approx 0.75$, which is very much at par with recent observations. Further, the present numerical values of the cosmological parameters obtained in the process are also quite absorbing, since the age of the universe ($13.86 < t_0 < 14.26$) Gyr, the present value of the Hubble parameter ($69.24 < H_0 < 69.96$) $\mathrm{Km.s^{-1}Mpc^{-1}}$, so that $0.991 < H_0t_0 < 1.01$ fit with the observation with appreciable precision. Numerical analysis also reveals that the state finder $\{r, s\} = \{1, 0\}$, which establishes the correspondence of the present model with the standard $\Lambda$CDM universe. Last but not the least important outcome is: considering the CMBR temperature at decoupling ($z \sim 1080$) to be $3000$, required for recombination, it's present value is found to be $2.7255$, which again fits the observation with extremely high precision. Thus, non-minimally coupled scalar-tensor theory of gravity appears to serve as a reasonably fair candidate for describing the evolution history of our observable universe, beyond quantum domain.\\

In the mean time new Planck's data is released \cite{Planck181, Planck182}, which imposed even tighter constraints on the inflationary parameters. In this manuscript, we therefore pose if the theory \cite{Beh} admits these new constraints. However, earlier we considered a particular form of coupling parameter along with the potential in the form $V(\phi) = V_0\phi^4 - B\phi^2$, where, $V_0$ and $B$ are constants \cite{Beh}. Here instead, we choose different forms of the coupling parameters and also different potentials to study the inflationary regime. In the following section 2, we describe the model, write down the field equations, find the parameters involved in the theory in view of a general conserved current. We also present the scalar-tensor equivalent form of the action in Einstein's frame to find the inflationary parameters. In section 3, we choose different forms of the coupling parameters and associated potentials to test the viability of the model in view of the latest released data from Planck \cite{Planck181, Planck182}. We conclude in section 4.

\section{The model, Conserved current, scalar-tensor equivalence and inflationary parameters:}

We start with the non-minimally coupled scalar-tensor theory of gravity, for which the action is expressed in the form,

\be \label{2.1} A = \int \left[f(\phi) R - {\omega(\phi)\over \phi}\phi_{,\mu}\phi^{^,\mu} - V(\phi) - \mathcal{L}_m \right]\sqrt{-g} d^4 x, \ee
where, $\mathcal{L}_m$ is the matter Lagrangian density, $f(\phi)$ is the coupling parameter, while, $\omega(\phi)$ is the variable Brans-Dicke parameter. The field equations are,

\be \label{2.2} \Big(R_{\mu\nu} - {1\over 2} g_{\mu\nu}R\Big) f(\phi) + g_{\mu\nu}\Box f(\phi)  - f_{;\mu;\nu} -{\omega(\phi)\over \phi}\phi_{,\mu}\phi_{,\nu} + {1\over 2} g_{\mu\nu}\Big(\phi_{,\alpha}\phi^{,\alpha} + V(\phi)\Big) = T_{\mu\nu},\ee
\be \label{2.3} R f' + 2 {\omega(\phi)\over \phi}\Box\phi +\Big({\omega'(\phi)\over \phi} -{\omega(\phi)\over \phi^2}\Big)\phi_{,\mu}\phi^{,\mu} - V'(\phi) = 0,\ee
where prime denotes derivative with respect to $\phi$, and $\Box$ denotes D'Alembertian, such that, $\Box f(\phi) = f''\phi_{,\mu}\phi^{^,\mu} - f'\Box\phi$. The model involves three parameters viz. the coupling parameter $f(\phi)$, the Brans-Dicke parameter $\omega(\phi)$ and the potential $V(\phi)$. It is customary to choose these parameters by hand in order to study the evolution of the universe. However, we have proposed a unique technique to relate the parameters in such a manner, that choosing one of these may fix the rest \cite{cons1, cons2, cons3, cons4, Beh}. This follows in view of a general conserved current which is admissible by the above pair of field equations, briefly enunciated below.\\

\noindent
The trace of the field equation \eqref{2.2} reads as,

\be\label{trace} R f - 3\Box f - {\omega(\phi)\over \phi}\phi_{,\mu}\phi^{^,\mu} -2V = T_\mu^\mu = T.\ee
Now eliminating the scalar curvature between equations \eqref{2.3} and \eqref{trace}, one obtains,

\be\label{combination1}  \Big(3f'^2 + {2\omega f\over \phi}\Big)' \phi_{,\mu}\phi^{^,\mu} +  \Big(3f'^2 + {2\omega f\over \phi}\Big)\Box\phi + 2 f'V - f V = f'T,\ee
which may then be expressed as,

\be\label{combination2}  \Big[\Big(3f'^2 + {2\omega f\over \phi}\Big)^{1\over 2}\phi^{;\mu}\Big]_{,\mu} - {f^3\over 2\Big(3f'^2 + {2\omega f\over \phi}\Big)^{1\over 2}}\Big({V\over f^2}\Big)' = {f'\over 2\Big(3f'^2 + {2\omega f\over \phi}\Big)^{1\over 2}} T,\ee
and finally as,

\be\label{2.4} \left(3f'^2 + {2\omega f\over \phi}\right)^{1/2}\left[\left(3f'^2 + {2\omega f\over \phi}\right)^{1/2}\phi^{~;\mu}\right]_{;\mu} -f^3\left({V\over f^2}\right)' = {f'\over 2} T^\mu_\mu.\ee
Thus there exists a conserved current $J^\mu$, where,

\be\label{2.5} J^\mu_{;\mu} = \left[\left(3f'^2 + {2\omega f\over \phi}\right)^{1/2}\phi^{~;\mu}\right]_{;\mu} = 0.\ee
for trace-less matter field ($T^\mu_\mu = T = 0$), provided

\be \label{2.6}V(\phi) \propto f(\phi)^2.\ee
To study cosmological consequence of such a conserved current, let us turn our attention to the minisuperspace model \eqref{1.1}, in which
the conserved current \eqref{2.5}, reads as

\be\label{2.8} \sqrt{\left(3f'^2 + {2\omega f\over \phi}\right)}a^3\dot\phi = C_1,\ee
in traceless vacuum dominated and also in radiation dominated eras. In the above, $C_1$ is the integration constant. Note that, fixing the form of the coupling parameter $f(\phi)$, the potential $V(\phi)$ is fixed in view of \eqref{2.6}, once and forever. Further, we use a relation \cite{Beh}

\be \label{2.9} 3 f'^2 + {2\omega f\over \phi} = \omega_0^2,\ee
where, $\omega_0$ is a constant, to fix the Brans-Dicke parameter as well. As a result, we obtain the relation

\be \label{2.10} a^3 \dot \phi = {C_1\over \omega_0} = C,\ee
$C$ being yet another constant. In the process, all the coupling parameters $f(\phi)$, $\omega(\phi)$ and the potential $V(\phi)$ may be fixed a-priori. Note that the above choice \eqref{2.9} finally leads to the conserved current associated with the canonical momenta conjugate to the scalar field $\phi$, in the absence of a variable Brans-Dicke parameter, i.e. for $\big({\omega(\phi)\over \phi} = {1\over 2}\big)$ with minimal coupling $\Big(f(\phi) = (16\pi G)^{-1} = {M_p^2\over 2}\Big)$. We shall work, in the typical unit, ${M_p^{2}\over 2} = c = 1$, and consider different forms of $f(\phi)$, that fixes the Brans-Dicke parameter $\omega(\phi)$ as well as the potential $V(\phi)$. In view of these known functional forms of the parameters of the theory, we focus our attention to study inflation, which must have occurred in the very early vacuum dominated universe. We relax the symmetry by adding an useful term in the potential $V(\phi)$, so that one of the terms act just as a constant in the effective potential. This ensures a constant value of the potential as the scalar field dies out, and this constant acts as an effective cosmological constant ($\Lambda_e$). In the process, the number of parameters increases to three ($\omega_0, V_0, V_1$), which is essential to administer good fit with observation.\\

\subsection{Scalar-tensor equivalence and inflationary parameters:}

As mentioned, it is convenient and hence customary to study inflationary evolution in the Einstein's frame under suitable transformation of variables, where possible. Therefore, in order to study inflation, we consider very early vacuum dominated ($p = 0 = \rho$, for which trace of the matter field identically vanishes and symmetry holds) era, and express the action \eqref{2.1} in the form,

\be \label{3.1} A = \int \left[f(\phi) R - {K(\phi)\over 2}\phi_{,\mu}\phi^{,\mu} - V(\phi)\right]\sqrt{-g}~ d^4 x,\ee
where, $K(\phi) = 2 {\omega(\phi)\over \phi}$. The above action \eqref{3.1} may be translated to the Einstein's frame under the conformal transformation ($g_{E\mu\nu} = f(\phi) g_{\mu\nu}$) to take the form \cite{Conformal},

\be \label{3.2} A = \int \left[R_E - {1\over 2}\sigma_{E,\mu}{\sigma_{E}}^{,\mu} - V_E(\sigma(\phi))\right]\sqrt{-g_E}~ d^4 x,\ee
where, the subscript $`E$' stands for Einstein's frame. The effective potential ($V_E$) and the field ($\sigma$) in the Einstein's frame may be found from the following expressions,

\be \label{3.3} V_E = {V(\phi)\over f^2(\phi)};\hspace{0.4 in}\mathrm{and,}\hspace{0.4 in}\left({d\sigma\over d\phi}\right)^2 = {K(\phi)\over f(\phi)} + 3 {f'^2(\phi)\over f^2(\phi)}= {2\omega(\phi)\over \phi f(\phi)} + 3 {f'^2(\phi)\over f^2(\phi)}.\ee
In view of the action (\ref{3.2}), it is also possible to cast the field equations, viz. the Klein-Gordon and the ($^0_0$) equations of Einstein as,

\be\label{FE}\begin{split}& \ddot\sigma +3{H}\dot\sigma + V_E' = 0;\hspace{0.6 in} 3{H}^2 = \frac{1}{2} \dot\sigma^2 + V_E,\end{split}\ee
where, ${H}={\dot a_E\over a_E}$ denotes the expansion rate, commonly known as the Hubble parameter. The slow-roll parameters and the number of e-foldings, then admit the following forms,

\be\label{3.4} \epsilon = \Big({V'_E\over V_E}\Big)^2\Big({d\sigma\over d\phi}\Big)^{-2}; \hspace{0.1 in} \eta = 2\left[\Big({V''_E\over V_E}\Big)\Big({d\sigma\over d\phi}\Big)^{-2} - \Big({V'_E\over V_E}\Big)\Big({d\sigma\over d\phi}\Big)^{-3}{d^2\sigma\over d\phi^2}\right];\hspace{0.1 in} N = \int_{t_i}^{t_f} H dt ={1\over 2\sqrt 2}\int_{\phi_e}^{\phi_b} {d\phi\over \sqrt \epsilon}{d\sigma\over d\phi},\ee
where, $t_i,~ t_f$ stand for the initiation time and the end time, while $\phi_b,~ \phi_e$ stand for the values of the scalar field at the beginning and at the end of inflation respectively. Comparing expression for the primordial curvature perturbation on super-Hubble scales produced by single-field inflation ($P_\zeta(k)$) with the primordial gravitational wave power spectrum ($P_t(k)$), one obtains the tensor-to-scalar ratio for single-field slow-roll inflation $r = {P_t(k)\over P_\zeta(k)} = 16\epsilon$, while, the scalar tilt, conventionally defined as $n_s-1$ may be expressed as $n_s - 1 = -6\epsilon + 2\eta$, or equivalently $n_s = 1 - 6\epsilon + 2\eta$, dubbed as scalar spectral index. According to the latest released results, the scalar to tensor ratio $r \le 0.16$ (TT,TE,EE+lowEB+lensing), while $r \le 0.07$ (TT,TE,EE+lowE+lensing+BK14+BAO) \cite{Planck181, Planck182}. Further, combination of all the data (TT+lowE, EE+lowE, TE+lowE, TT,TE,EE+lowE, TT,TE,EE+lowE+lensing) constrain the scalar spectral index to $0.9569 \le n_s \le 0.9815$ \cite{Planck181, Planck182}. It is useful to emphasize that under the present choice of unit ${M_p^2\over 2} = c = 1$ (which although appears to be a bit unusual but doesn't cause any harm), $\phi$ controls the cosmological evolution in the manner $\phi > 1$ corresponds to the inflationary stage, $\phi \sim 1$ describes the end of inflation while $\phi < 1$ is the low energy regime which triggers matter dominated era.

\section{Inflation with power law and exponential potentials:}

In the non-minimal theory, the flat section of the potential $V(\phi)$ responsible for slow-roll is usually distorted. However, flat potential is still obtainable if the Einstein's frame potential $V_E$ is asymptotically constant \cite{PY1, PY2}. Note that, the symmetry explored in section 2, makes the potential ($V_E$) in the Einstein's frame to be constant, once and forever. Thus, we need to relax the symmetry as required by the condition \eqref{2.6}, by taking into account additional term in the potential, viz. a constant term $V_0$, or even a functional form, to ensure that the effective potential in the Einstein's frame ($V_E$) is asymptotically constant. At the end of inflation, the universe becomes cool due to sudden large (exponential, in the present case) expansion. Therefore, in order that the structure we live in are formed, the universe must be reheated and take the state of a hot thick soup of plasma (the so called hot Big-Bang). This phenomena is possible if at the end of inflation, the scalar field starts oscillating rapidly on the Hubble time scale, about the minimum of the potential. In the process, particles are created under standard quantum field theoretic (in curved space-time) approach, which results in the re-heating of the universe. The universe then eventually transits to the radiation dominated era. At that epoch, the additional term may be absorbed in the potential if it is a constant term ($V_0$), or may even be neglected in case it is a function (since $\phi$ goes below the Planck's mass), without any loss of generality, to reassure symmetry. The symmetry leads to the first integral of certain combination of the field equations, which helps in solving the field equations leading to a Friedmann-like radiation dominated era, as shown earlier \cite{Beh}. However, in the present manuscript, we only concentrate upon inflationary regime and of-course study possibility of graceful exit from inflation. In the following subsection, we shall study different power law potentials, while in the next we shall deal with exponential potential. We consider de-Sitter solution in the form $a \propto e^{Ht}$, where the Hubble parameter $H$ is slowly varying during inflation. We repeat that according to our current choice of units (${M_p^2\over 2} = 1$), which although is uncommon, but doesn't create any problem whatsoever, the value of $f(\phi)$ at the end of inflation must be a little greater than $1$.

\subsection{Power law potential $f(\phi) = \phi^n$:}

Under the choice, $f(\phi) = \phi^{n}$, the potential is $V(\phi) \propto \phi^{2n}$. We shall take into account three different values of $n$, viz. $n = 1, {3\over 2}~ \mathrm{and}~2$, in the following three sub-subsections. For each value of $n$ we shall study different cases taking into account different additive terms. In the first place however, we shall consider an additive constant $V_0$ in all the three cases, viz.

\be\label{V}  \mathrm{Case-1:} ~V(\phi) = V_1 \phi^{2n} + V_0, \ee
corresponding to which, one can now find the expression for the Brans-Dicke parameter $\omega(\phi)$, the potential $V_E(\sigma)$ in the Einstein's frame, the expression for ${d\sigma\over d\phi}$, and the slow-roll parameters $\epsilon, ~\eta$ along with the number of e-foldings $N$, in view of the equations \eqref{2.9}, \eqref{3.3} and \eqref{3.4} respectively as,

\be \mathrm{Case-1:}\Bigg\{~\label{B-D}\begin{split} & \omega(\phi) = \frac{\omega_0^2 - 3n^2\phi^{2(n - 1)}}{2\phi^{(n - 1)}},\hspace{0.3 in} V_E = V_1 + V_0\phi^{-2n}, \hspace{0.65 in} \left({d\sigma\over d\phi}\right)^2 = {\omega_0^2\over \phi^{2n}},\\&
\epsilon = {4n^2 V_0^2\phi^{2(n-1)}\over \omega_0^2(V_0 + V_1\phi^{2n})^2},\hspace{0.5 in}  \eta = {4n(n+1) V_0\phi^{2(n-1)}\over \omega_0^2(V_0 + V_1\phi^{2n})},\hspace{0.35 in} N = {\omega_0^2\over 4\sqrt 2n V_0}\int_{\phi_e}^{\phi_b}{V_0 + V_1 \phi^{2n}\over \phi^{2n-1}} d\phi.\end{split}\ee
The effect of the constant term $V_0$ is now clearly noticeable, since when $\phi$ is large, second term in the Einstein's frame potential $(V_E)$ becomes insignificantly small, for $n \ge 1$, and it almost becomes (non-zero) constant, assuring slow-roll. On the contrary, if $V_0$ is set to vanish from the very beginning, the Einstein's frame potential $V_E = V_1$ would remain flat always, and the universe would have been ever-inflating.\\

\noindent
We shall also consider a functional additive term in the potential for all the cases under consideration, such the the potential reads as

\be\label{V1}  \mathrm{Case-2:} ~V(\phi) = V_1 \phi^{2n} + V_0 \phi^m. \ee
In view of the above potential \eqref{V1} it is possible to find the expression for the Brans-Dicke parameter $\omega(\phi)$, the potential $V_E(\sigma)$ in the Einstein's frame, the expression for ${d\sigma\over d\phi}$, and the slow-roll parameters $\epsilon, ~\eta$ along with the number of e-foldings $N$, in view of the equations \eqref{2.9}, \eqref{3.3} and \eqref{3.4} respectively as,

\be \mathrm{Case-2:}\Bigg\{~\label{B-D1}\begin{split} & \omega(\phi) = \frac{\omega_0^2 - 3n^2\phi^{2(n - 1)}}{2\phi^{(n - 1)}},\hspace{0.1 in} V_E = V_1 + V_0\phi^{m-2n}, \hspace{0.1 in} \left({d\sigma\over d\phi}\right)^2 = {\omega_0^2\over \phi^{2n}},\hspace{0.1 in}
\epsilon = {(m-2n)^2 V_0^2\phi^{2(m-n-1)}\over \omega_0^2[V_1 + V_0\phi^{(m-2n)}]^2},\\&  \eta = {2V_0(m-2n)(m-n-1)\phi^{(m-2)}\over \omega_0^2[V_1 + V_0\phi^{(m-2n)}]}, \hspace{0.5 in} N = {\omega_0^2\over 2\sqrt 2 (m-2n) V_0}\int_{\phi_e}^{\phi_b}{V_1 + V_0 \phi^{m-2n}\over \phi^{(m-1)}} d\phi.\end{split}\ee

\noindent
In the following sub-subsections, we shall take three different values of $n$, as already mentioned, and present the data set in tabular form along with appropriate plots, to demonstrate the behaviour of the slow-roll parameters in comparison with the latest data set released by Planck \cite{Planck181, Planck182}. Different additive terms, as indicated, will be considered in each subcase separately. In the subsection (3.1.3), we shall consider an additional case with a pair of additive terms in the form of a whole square.\\

\subsubsection{$n=1,~ f(\phi) = \phi$.}

\textbf{Case-1:}
Under the choice $n = 1$, the potential \eqref{V} takes the form $V(\phi) = V_0 + V_1\phi^2$, and thus the parameters of the theory under consideration \eqref{B-D} read as,

\be \label{Para1}\begin{split}& \omega(\phi) = \frac{\omega_0^2 - 3}{2},~~~{d\sigma\over d\phi}={\omega_0\over \phi},\hspace{0.3 in} V_E = V_1 + V_0\phi^{-2},\hspace{0.3 in}\epsilon ={4 V_0^2\over \omega_0^2(V_0 + V_1\phi^2)^2},\\&
\eta = {8V_0\over \omega_0^2(V_0 + V_1\phi^2)},\hspace{0.3 in}
N = \frac{\omega_0^2}{4\sqrt 2V_0}\left[V_1\left({\phi_b^2\over 2} - {\phi_e^2\over 2}\right) + V_0(\ln{\phi_b} - \ln{\phi_e}) \right].\end{split}\ee
In view of the above forms of the slow roll parameters \eqref{Para1}, we present table-1 and table-2, underneath, corresponding to two different values of the parameter $V_1 > 0$. The wonderful fit with the latest data sets released by Planck \cite{Planck181, Planck182} is appreciable particularly because $0.968 \approx n_s < 0.982$, while $r <0.0278$. Further, the number of e-fold ($36 \le N \le 62$) is sufficient to alleviate the horizon and flatness problems. Figure 1 and figure 2 are the two plots $r$ versus $n_s$ and $r$ versus $\omega_0$ respectively, presented for visualization. For example, the figures clearly depict that the plot which represents data sets corresponding to table 2, appears to be even better.\\

\begin{figure}[h!]
\begin{minipage}[h]{0.47\textwidth}
      \centering
\begin{tabular}{|c|c|c|c|c|c|}
\hline\hline
 $\omega_0$ & $|\eta|$ & $r=16\epsilon$ & $\phi_e$ & $n_s$ & $N$ \\\hline
 16.0&.010418 & .0278 & 1.060 &.9688 & 36 \\\hline
 16.5&.009795 & .0261 & 1.059  &.9706 & 38\\\hline
 17.0&.009227 & .0246 & 1.057 &.9723 & 41 \\\hline
 17.5&.008707& .0232 & 1.056 &.9739 & 43 \\\hline
 18.0&.008230& .0219 & 1.054 &.9753 & 46 \\\hline
 18.5&.007792& .0208 & 1.053  &.9766 & 48 \\\hline
 19.0&.007387& .0197 & 1.051  &.9778 & 51 \\\hline
 19.5&.007013& .0187 & 1.050 &.9790 & 54 \\\hline
 20.0&.006667& .0178& 1.049 &.9800 & 56 \\\hline
 20.5&.006345& .0169 & 1.048 &.9810 & 59 \\\hline
 21.0&.006047& .0161 & 1.047 &.9819 & 62 \\\hline
\hline
\end{tabular}
      \captionof{table}{$f(\phi)=\phi$, (case-1): ${\phi_b}=2.0$\\${V_0}=-0.9\times {10^{-13}\mathrm{T}^{-2}},~{V_1}=0.9\times {10^{-13}\mathrm{T}^{-2}}.$}
      \label{table:1}
   \end{minipage}%
\hfill%
\begin{minipage}[h]{0.47\textwidth}
\centering
\begin{tabular}{|c|c|c|c|c|c|c|}
  \hline\hline
 $\omega_0$ &$|\eta|$& $r=16\epsilon$ & $\phi_e~$ &$n_s$& $N$ \\\hline
 14.5&.011047 &.0257 & 1.012 &.9682 & 36 \\\hline
 15.0&.01032 &.0239 & 1.009  & .9703 & 38 \\\hline
 15.5&.009667&.0225 & 1.008 &.9722& 41 \\\hline
 16.0&.009072 &.0210 & 1.006 & .9739 & 43 \\\hline
 16.5&.008531 &.0198 & 1.005 &.9755& 46 \\\hline
 17.0&.008037 & .0187 &1.003 &.9769 & 49 \\\hline
 17.5&.007584 &.0176 & 1.001 & .9782 & 52 \\\hline
 18.0 &.007168 &.0166 & 1.000 &.9794 & 55 \\\hline
 18.5 &.006786 &.0158 & 0.9986 &.9805 & 59 \\\hline
 19.0 &.006433 &.0149 & 0.9973 &.9815 & 62 \\\hline
 \hline
\end{tabular}
\captionof{table}{$f(\phi) = \phi$, (case-1): ${\phi_b}=2.0$.\\${V_0}=-0.9\times {10^{-13}\mathrm{T}^{-2}},~{V_1}=1.0\times{10^{-13}\mathrm{T}^{-2}}~.$}
      \label{table:2}
   \end{minipage}%
\end{figure}

\begin{figure}
\begin{minipage}[h]{0.47\textwidth}
\centering
\includegraphics[ width=0.9\textwidth] {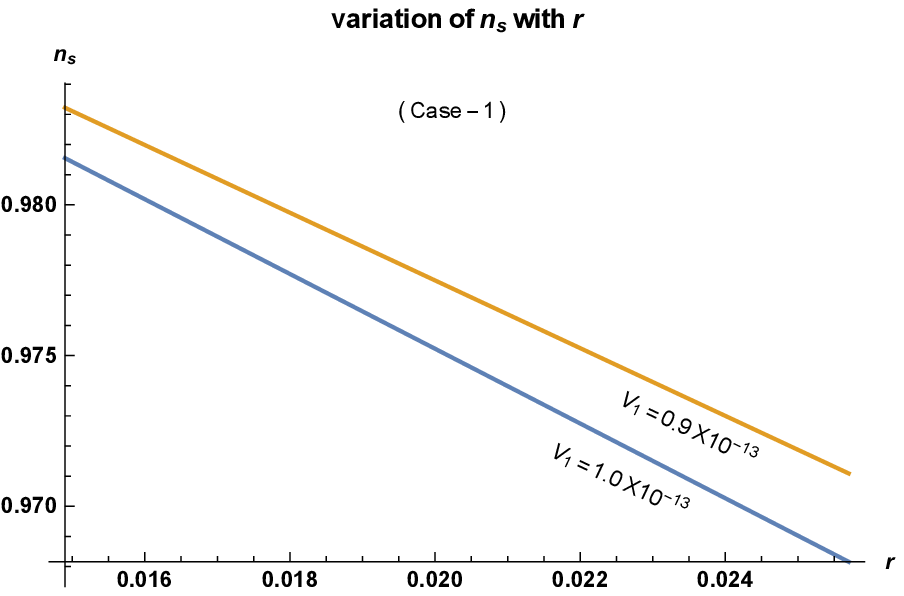}
 \caption{$f = \phi$, (case-1): Yellow ochre  and blue colours represent table-1 and table-2 data respectively.}
      \label{fig:1}
   \end{minipage}%
\hfill%
\begin{minipage}[h]{0.47\textwidth}
\centering
\includegraphics[ width=0.9\textwidth] {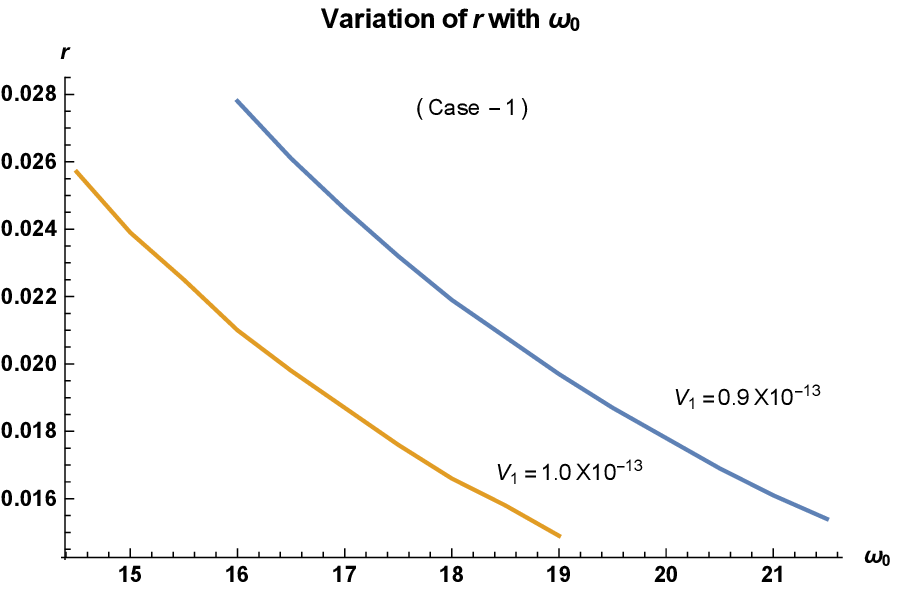}
 \caption{$f = \phi$, (case-1): Blue and yellow ochre colours represent table-1 and table-2 data respectively.}
      \label{fig:2}
   \end{minipage}%
\end{figure}

\noindent
One very interesting feature is that the above data sets remain unaltered even if the sign of $V_0$ and $V_1$ are interchanged. Note that, second derivative of the potential has to be positive, since it represents effective mass of the scalar field. In view of the forms of the potentials $V(\phi)$ and $V_E(\sigma)$ presented in \eqref{V} and \eqref{B-D} the effective mass of the scalar fields $\phi$ and $\sigma$ respectively are,

\be {d^2V\over d\phi^2} = 2V_1;~~~{d^2V_E\over d\sigma^2} = 6{V_0 \over \phi^4}.\ee
In our data set, we keep $V_1 > 0$, since $\phi$ is the scalar field under consideration, while translation to $\sigma$ only amounts to handling the situation with considerable ease. However, as a matter of taste if one favours Einstein's frame over Jordan's frame, it is possible to revert the sign and keep $V_0 > 0$,  without changing the data set.\\

\noindent
As mentioned, at the end of inflation, the scalar field must oscillate rapidly so that particles are produced and the universe turns to the phase of: a hot thick soup of plasma, commonly called the `hot big-bang'. This phenomena is dubbed as graceful exit, which is required for the structure formation together with the formation of CMB. We therefore proceed to check if the present model admits graceful exit from inflation. Here, $V_E=V_1+{V_0\over\phi^2}$, and so one can express (\ref{FE}) as,

\be\label{n10}\begin{split}& {3{H}^2\over V_1} = \frac{\dot\sigma^2}{2V_1} +\left(1+{ V_0\over {V_1\phi^2}}\right).\end{split}\ee
At large value of the scalar field, which in the present unit $\phi > 1$,  we obtained slow-roll. However, as the scalar field falls below the Planck's mass $M_p$, then the Hubble rate ${H}$ also decreases, and once it falls below the effective mass $V_1$ i.e. ${H} \ll V_1 $, then the above equation may be approximated to,

\be {\dot\sigma^2}=2i^2\left(V_1+{V_0\over\phi^2}\right) \hspace{0.3 in}\longrightarrow\hspace{0.3 in}{\dot\phi}=i{\phi\over\omega_0}\sqrt{2\left(V_1+{V_0\over\phi^2}\right)},\ee
Where, $\dot\sigma={\omega_0\over \phi}\dot\phi$, in view of \eqref{Para1}. Thus, finally we get,

\be \phi = {1\over 2V_1}\left[(1-V_0V_1) \cos{\left({\sqrt {2V_1}\over \omega_0}t\right)}+ i (1+ V_0V_1)\sin{\left({\sqrt {2V_1}\over \omega_0}t\right)}\right],\ee
which is an oscillatory solution, and the field then oscillates many times over a Hubble time. This coherent oscillating field corresponds to a condensate of non-relativistic massive (inflaton) particles, which ensures graceful exit from the inflationary regime, driving a matter-dominated era at the end of inflation. There is a long standing debate regarding the physical frame. It appears that most of the people favour Einstein's frame over Jordan's frame (we have briefly discussed the issue in conclusion). In this regard, it is important to mention that since in view of \eqref{Para1} $\sigma = \omega_0\ln{\phi}$, therefore $\sigma$ executes oscillatory behaviour as well. \\

\noindent
\textbf{Case-2:} Under the same situation $f(\phi) = \phi$, let us now consider, $V(\phi) = V_1 \phi^{2} + V_0\phi^{4}$, where instead of a constant term, we have added a quartic term in the potential. The expression for the Brans-Dicke parameter ($\omega(\phi)$), the potential ($V_E$) in the Einstein's frame, ${d\sigma\over d\phi}$, the slow-roll parameters $\epsilon, ~\eta$ and the number of e-foldings $N$, may then be found in view of the equation \eqref{B-D1}, respectively as,

\be \label{ParaN}\begin{split}& \omega(\phi) = \frac{\omega_0^2 - 3}{2\phi},~~~{d\sigma\over d\phi}={\omega_0\over \phi},\hspace{0.3 in} V_E = V_1+V_0\phi^{2},\hspace{0.3 in}\epsilon ={4V_0^2\phi^4\over \omega_0^2(V_1+V_0\phi^2)^2},\\&
\eta = {8 V_0\phi^2\over \omega_0^2(V_1+V_0\phi^2)},\hspace{0.3 in}
N =\int_{\phi_e}^{\phi_b}\frac{\omega_0^2(V_1+V_0\phi^2)}{4\sqrt 2V_0\phi^3} d\phi .\end{split}\ee
Although, the potential $V_E$ does not appear to attend a flat section, the smallness  of the value of $\eta$ confirms that there indeed exists a flat section, admitting slow-roll. In fact, in the Einstein's frame \eqref{3.2}, this is just the case of a standard inflation field theory with quadratic potential. Followings tables 3 and 4, for $V_1 > 0$, together with the associated plots $n_s$ versus $r$ and $r$ versus $\omega_0$  here again depict appreciably good fit with the recent released Planck's data set, particularly because $0.97 \leq n_s \leq 0.98$ while $r \le 0.098$. The figure 3 depicts that data of table-3 is somewhat better.
\begin{figure}[h!]
\begin{minipage}[h]{0.47\textwidth}
      \centering
\begin{tabular}{|c|c|c|c|c|c|}
 \hline\hline
 ${\omega_0}$&$|\eta|$ & $r=16\epsilon$ & ${\phi_e} ~$ & $n_s$& $N$ \\\hline
 114&.002607 &.08834 &1.0581 & .9721 & 38 \\\hline
 116 &.002518 &.08532 &1.0580 & .9730 & 39 \\\hline
 118 &.002433 &.08245 &1.0578 & .9739 & 41 \\\hline
 120 &.002353 &.07972 & 1.0577 &.9748 &42 \\\hline
 122 &.002276 &.07713 & 1.0575 &.9756 & 44 \\\hline
 124&.002204 &.07466 &1.0574 &.9764 & 45 \\\hline
 126 &.002134 &.07231 & 1.0572 &.9772 & 46\\\hline
 128 &.002068 &.07007 &1.0571 &.9779 & 48\\\hline
 130 &.002005 &.06793 & 1.0570 &.9785 & 49 \\\hline
 132 &.001945 &.06589 & 1.0569 &.9792 & 51 \\\hline
 134 &.001887 &.06393 &1.0567 &.9798 & 53 \\\hline
 136 &.001832 &.06207 & 1.0566 &.9804 & 54 \\\hline
 138 &.001780 &.06028 & 1.0565 &.9810 & 56 \\\hline
 140 &.001729 &.05857 & 1.0564 &.9815 & 57 \\\hline
 \hline
\end{tabular}
\captionof{table}{$f(\phi) = \phi$, (case-2): ${\phi_b}=1.2$,\\ ${V_0}=-1.0\times {10^{-20}\mathrm{T}^{-2}};{V_1}=1.1\times {10^{-20}\mathrm{T}^{-2}}.$}
      \label{table:3}
   \end{minipage}%
\hfill%
\begin{minipage}[h]{0.47\textwidth}
      \centering
\begin{tabular}{|c|c|c|c|c|c|}
 \hline\hline
 ${\omega_0}$&$|\eta|$ & $r=16\epsilon$ & ${\phi_e} ~$ & $n_s$& $N$ \\\hline
 84& .003711&.09715 & 1.0121 & .9710 & 36 \\\hline
 86 &.003540&.09268 &1.0118 & .9723 & 38 \\\hline
 88 &.003381&.08852 &1.0115 & .9736 & 40 \\\hline
 90 &.003232&.08463 &1.0113 & .9747 & 42 \\\hline
 92& .003093&.08099 & 1.0110 &.9758 & 44 \\\hline
 94 &.002963&.07758 &1.0108 & .9768 & 46 \\\hline
 96 &.002841&.07438 & 1.0105 &.9778 & 47\\\hline
 98 &.002726&.07138 & 1.0103 &.9787 & 49\\\hline
 100 &.002618&.06855 & 1.0101 &.9795 & 51 \\\hline
 102 &.002517&.06589 & 1.0099 &.9803 & 54 \\\hline
 104 &.002421&.0634 & 1.0097 &.9810 & 56 \\\hline
 106 &.002330&.06101 & 1.0096 &.9818 & 58 \\\hline
 \hline
\end{tabular}
\captionof{table}{$f(\phi) = \phi$, (case-2): ${\phi_b}=1.2$, \\${V_0}=-1.0\times {10^{-20}\mathrm{T}^{-2}};{V_1}=1.0\times {10^{-20}\mathrm{T}^{-2}}.$}
      \label{table:4}
   \end{minipage}%
\end{figure}
\begin{figure}
\begin{minipage}[h]{0.47\textwidth}
\centering
\includegraphics[ width=0.9\textwidth] {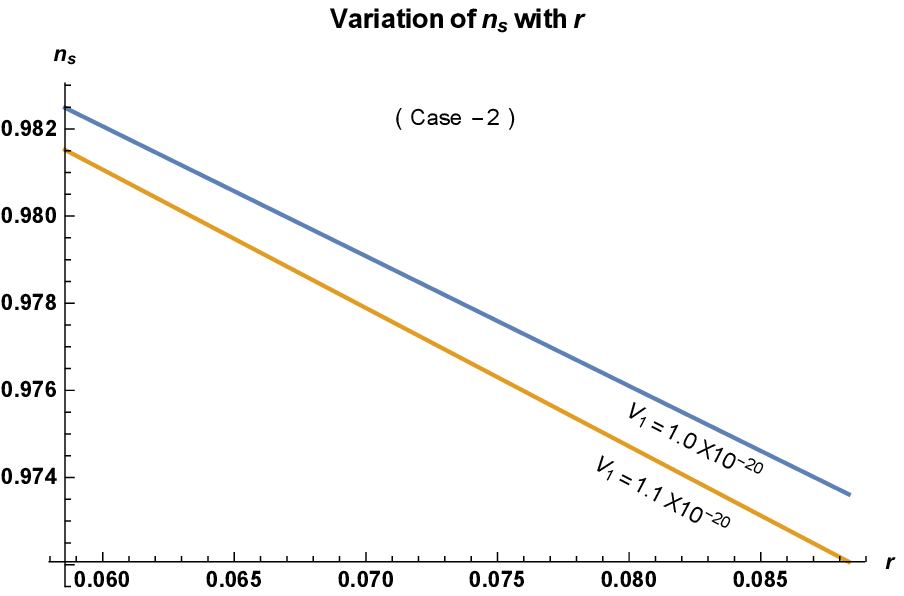}
 \caption{$f = \phi$. Yellow ochre and blue colours represent table 3 and 4 respectively.}
      \label{fig:3}
   \end{minipage}%
\hfill%
\begin{minipage}[h]{0.47\textwidth}
\centering
\includegraphics[ width=0.9\textwidth] {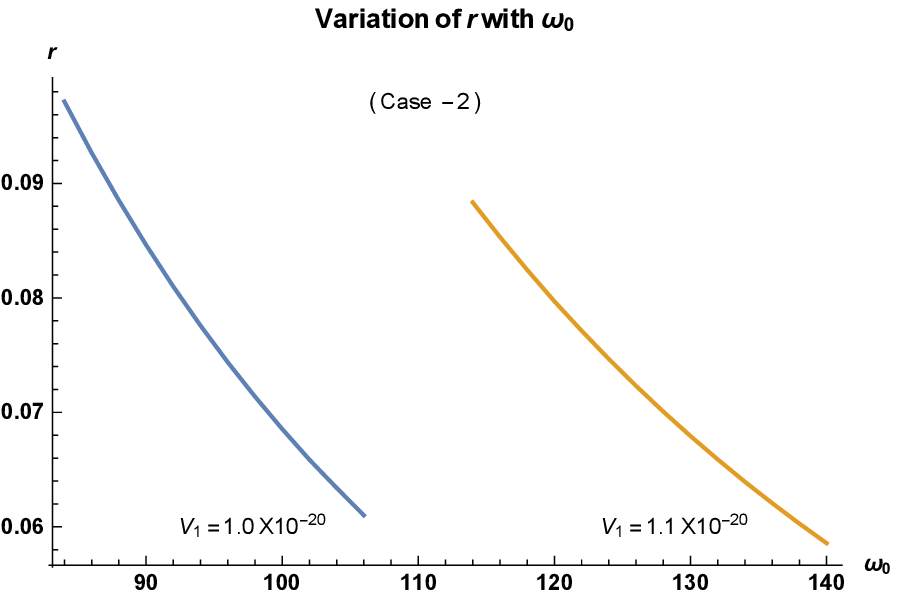}
 \caption{$f = \phi$. Yellow ochre and blue colours represent table 3 and 4 respectively, unlike previous cases.}
      \label{fig:4}
   \end{minipage}%
\end{figure}
As before here again we test if the model associated with a different potential admits graceful exit. Here $V_E=(V_1+{V_0\phi^2})$, and so from (\ref{FE}) one obtains,

\be\label{n11}\begin{split}& {3{H}^2\over V_1} = \frac{\dot\sigma^2}{2V_1} +\left(1+{V_0\over V_1}{\phi^2}\right)\end{split}\ee
As the Hubble rate ${H}\ll V_1$, the above equation can be approximated as, ${\dot\sigma^2}=2i^2(V_1+{V_0\phi^2})$, yields ${\dot\phi}=i{\phi\over\omega_0}\sqrt{2(V_1+{V_0\phi^2})}$, where, $\dot\sigma=\dot\phi{\omega_0\over \phi}$. Finally we get,

\be \phi = \frac{2V_1\left[(1-V_0V_1) \cos{\left({\sqrt {2V_1}\over \omega_0}t\right)}+ i (1+ V_0V_1)\sin{\left({\sqrt {2V_1}\over \omega_0}t\right)}\right]}{1 + V_0^2V_1^2 - 2V_0V_1 \cos{\left({2\sqrt{2V_1}\over \omega_0}t\right)}},\ee
The oscillatory behaviour of the scalar field clearly ensures graceful exit from inflationary regime, as already discussed, and in view of \eqref{ParaN} $\sigma$ also executes oscillatory behaviour.

\subsubsection{$n={3\over 2}, ~f(\phi) = \phi^{3\over 2}$.}

\noindent
\textbf{Case-1:}\\

\noindent
Under the choice $n = {3\over 2}$, $f(\phi) = \phi^{3\over 2}$, and the potential \eqref{V} takes the cubic form, $V(\phi) = V_0 + V_1 \phi^3$. Thus the expressions for the parameters of the theory under consideration along with the slow-roll parameters \eqref{B-D} are,

\be \label{Para3b2}\begin{split}& \omega(\phi) = \frac{4\omega_0^2 - 27\phi}{8\sqrt \phi},~~~{d\sigma\over d\phi}={\omega_0\over \phi^{3\over 2}};\hspace{0.3 in} V_E = V_1 + V_0\phi^{-3},\hspace{0.3 in}\epsilon ={9 V_0^2\phi\over \omega_0^2(V_0 + V_1\phi^3)^2},\\&
\eta = {15 V_0\phi\over \omega_0^2(V_0 + V_1\phi^3)},\hspace{0.3 in}
N = \frac{\omega_0^2}{6\sqrt 2V_0}\left[V_0\left({1\over \phi_e} - {1\over \phi_b}\right) + {V_1\over 2}(\phi_b^2 - \phi_e^2) \right].\end{split}\ee
As before, we present two sets of data in tables 5 and 6, for two different values of $V_1 > 0$. Plots 5 and 6 depict the variations of the spectral index $n_s$ with the scalar-tensor ratio $r$ and the scalar-tensor ratio $r$ with the Brans-Dicke parameter $\omega_0$ respectively. Here again we observe that $r \le 0.0278$, and $0.96 < n_s < 0.9815$, which are very much within the stipulated observational range \cite{Planck181, Planck182}, while number of e-folding $N$ is sufficient to remove the flatness and the horizon problems.\\

\begin{figure}[h!]
\begin{minipage}[h]{0.47\textwidth}
      \centering
\begin{tabular}{|c|c|c|c|c|c|}
  \hline\hline
  $\omega_0~$&$|\eta|$ & $r=16\epsilon$ & $\phi_e~$ & $n_s$& $N$ \\\hline
  24 &.011314 &.0278 & 1.0408 & .9670 & 36 \\\hline
  25 &.010427 &.0256 & 1.0392 & .9696 & 39 \\\hline
  26 &.009640& .0237 & 1.0377 & .9719 & 42\\\hline
  27 &.008939 &.0219 & 1.0364 & .9739 & 46\\\hline
  28 &.008312 &.0204 & 1.0351 & .9757 & 49 \\\hline
  29 &.007749 &.0190 & 1.0339 & .9773 & 52 \\\hline
  30 &.007241& .0178 & 1.0328 & .9788 & 56 \\\hline
  31 &.006781& .0166 & 1.0318 & .9802 & 60 \\\hline
  32 &.006364& .0156 & 1.0308 & .9814 &64 \\\hline
  \hline
\end{tabular}
\captionof{table}{$f(\phi) = \phi^{3\over 2}$, (case-1): ${\phi_b}=1.7$,\\ ${V_0}=-0.9\times {10^{-13}\mathrm{T}^{-2}},~{V_1}=0.9\times {10^{-13}\mathrm{T}^{-2}}.$}
      \label{table:5}
   \end{minipage}%
\hfill%
\begin{minipage}[h]{0.47\textwidth}
      \centering
\begin{tabular}{|c|c|c|c|c|c|}
  \hline\hline
  $\omega_0~$&$|\eta|$ & $r=16\epsilon$ & $\phi_e~$ & $n_s$& $N$ \\\hline
  22.0 &.011816 &.0254 & 1.0076& .9668 & 36 \\\hline
  22.5 &.011217 &.0243 & 1.0067& .9683 & 38 \\\hline
  23.0 &.010811&.0233 & 1.0059& .9697& 39\\\hline
  23.5 &.010356 &.0223 & 1.0050& .9709 & 41 \\\hline
  24.0 &.009928 &.0214 & 1.0042& .9721 & 43 \\\hline
  24.5 &.009528 &.0205 & 1.0034& .9733 & 45 \\\hline
  25.0 &.009150 &.0197 & 1.0027& .9743 & 47\\\hline
  25.5 &.008795 &.0189 & 1.0019& .9753 & 48 \\\hline
  26.0 &.008460& .0182 & 1.0013 & .9762 & 50\\\hline
  26.5 &.008143 &.0175 & 1.0006& .9771 & 52 \\\hline
  27.0 &.007845 &.0169 & 1.000 & .9780 & 54 \\\hline
 \hline
\end{tabular}
\captionof{table}{$f(\phi) = \phi^{3\over 2}$, (case-1): ${\phi_b}=1.7$,\\${V_0}=-0.9\times {10^{-13}\mathrm{T}^{-2}},~{V_1}=1.0\times {10^{-13}\mathrm{T}^{-2}}.$}
      \label{table:6}
   \end{minipage}%
\end{figure}

\begin{figure}
\begin{minipage}[h]{0.47\textwidth}
\centering
\includegraphics[ width=0.9\textwidth] {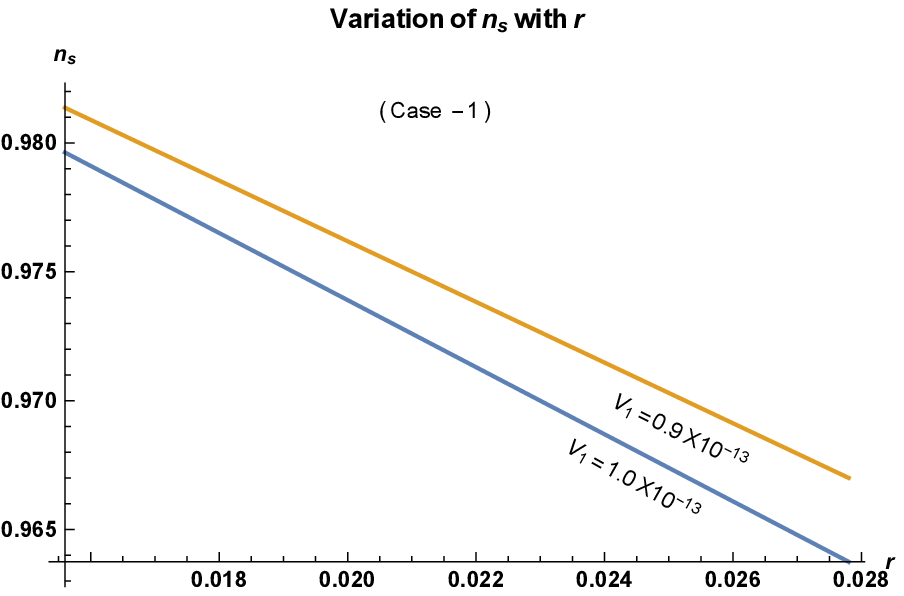}
 \caption{$f(\phi) = \phi^{3\over 2}$. Yellow ochre and blue colours represent table 5 and 6 respectively.}
      \label{fig:5}
   \end{minipage}%
\hfill%
\begin{minipage}[h]{0.47\textwidth}
\centering
\includegraphics[ width=0.9\textwidth] {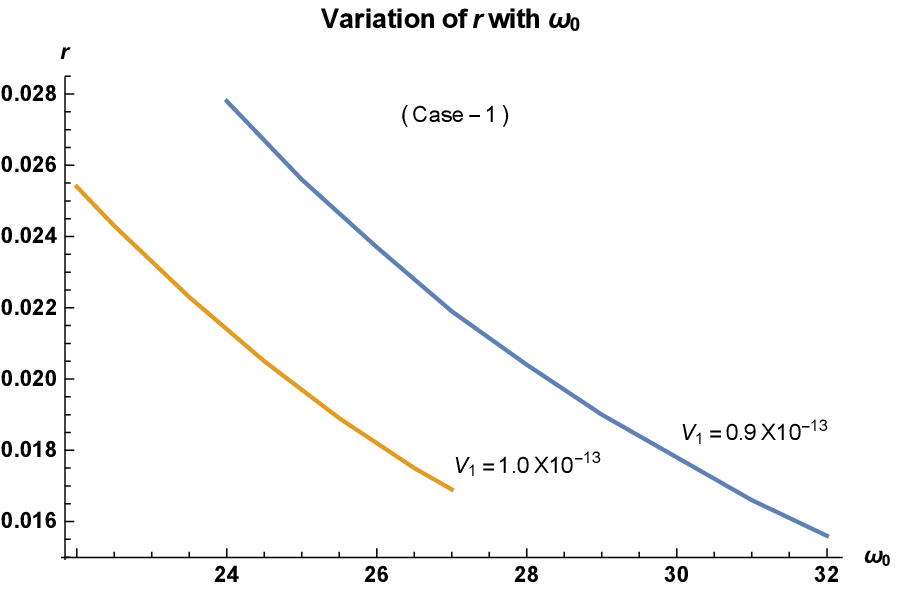}
 \caption{$f(\phi) = \phi^{3\over 2}$. Blue and yellow ochre colours represent table 5 and 6 respectively.}
      \label{fig:6}
   \end{minipage}%
\end{figure}

\noindent
To check the behaviour of the scalar field, we proceed as before, to find,
\be {3H^2\over V_1} = {\omega_0^2 \dot \phi^2\over2 V_1 \phi^{3}} + {V_0\over V_1\phi^3} + 1 \Longrightarrow \dot\phi^2 = -2{(V_0+V_1\phi^{3})\over \omega_0^2},\ee
under approximation, as the Hubble rate ${H}\ll V_1$, and using the relation $\dot \sigma = \omega_0{\dot\phi\over \phi^{3\over 2}}$, in view of \eqref{Para3b2}. The solution reads as,

\be\begin{split} \phi (t) = &\mathrm{InverseFunction}\Bigg[{2i\over \sqrt[4]{3} \sqrt[3]{V_1}\sqrt{V_0-\mathrm{\#}1^3 V_1}}\mathrm{Elliptic F}\Bigg[\mathrm{Sin}^{-1}\Bigg({1\over \sqrt[4]3}{\sqrt{-(-1)^{5\over 6} -i\#1 \sqrt[3]{{V_1}\over {V_0}}}}\Bigg), \sqrt[3]{-1}\Bigg]\\&
\left.\sqrt[3]{V_0} \sqrt{(-1)^{5/6} \left(\#1 \sqrt[3]{\frac{V_1}{V_0}}-1\right)} \sqrt{\mathrm{\#} 1^2 \sqrt[3]{\frac {V_1^2}{V_0^2}}+\# 1 \sqrt[3]{\frac{V_1}{V_0}}+1}\&\right]\left[c_1-t\right].
\end{split}\ee
In the above the hash tag ($\# n$) denotes $n$th argument of a pure function, and $c_1$ is a constant. Although, the solution is not obtainable in closed form, rather is a complicated inverse elliptic function, nevertheless its oscillatory behaviour is quite apparent, and $\sigma = -2{\omega_0\over \sqrt \phi}$ also oscillates as well.\\

\noindent \textbf{Case-2:} Cubic potentials with additive term have important consequence. For example, a potential in the form $V = {1\over 2} m \omega^2 x^2 - {1\over 3} bx^3$ can be used to model decay of metastable states \cite{321}, and it also describes the global flow \cite{322}. Further, the tunnelling rate in real time in the semiclassical limit may be found for arbitrary energy levels, while it's ground state agrees well with the result found by the instanton method \cite{323}. It is therefore worth to continue the present study in view of such an additive form in the cubic potential.\\

\noindent
Under the choice $n = {3\over 2}$, $f(\phi) = \phi^{3\over 2}$, and taking the potential as cubic form added with a quadratic term , i,e, $V(\phi) = V_1 \phi^3+V_0\phi^2 $, the expressions for the parameters of the theory under consideration along with the slow-roll parameters \eqref{B-D1} are,

\be \label{ParaN3b2}\begin{split}& \omega(\phi) = \frac{4\omega_0^2 - 27\phi}{8\sqrt \phi},~~~{d\sigma\over d\phi}={\omega_0\over \phi^{3\over 2}};\hspace{0.3 in} V_E = V_1 +{ V_0\over \phi},\hspace{0.3 in}\epsilon ={ V_0^2\phi\over \omega_0^2(V_0 + V_1\phi)^2},\\&
\eta = { V_0\phi\over \omega_0^2(V_0 + V_1\phi)},\hspace{0.3 in}
N = \frac{\omega_0^2}{2\sqrt 2V_0}\left[V_0\left({1\over \phi_e} - {1\over \phi_b}\right) + V_1\ln(\phi_b - \phi_e) \right].\end{split}\ee
We present two sets of data in tables 7 and 8 underneath, for two different values of $V_1 > 0$. Plots 7 and 8 depict the variations of the spectral index $n_s$ with the scalar-tensor ratio $r$ and the scalar-tensor ratio $r$ with the Brans-Dicke parameter $\omega_0$ respectively. Here again we observe that $r \le 0.062$, and $0.973 < n_s < 0.983$, which are again in excellent agreement of Planck's data \cite{Planck181, Planck182}, while the number of e-folding $N$ is also sufficient to remove the flatness and the horizon problems. It is interesting to note that the variation $n_s$ with $r$ for the two sets of data almost overlap in figure-7.
\begin{figure}[h!]
\begin{minipage}[h]{0.47\textwidth}
      \centering
\begin{tabular}{|c|c|c|c|c|c|}
  \hline\hline
  $\omega_0~$&$|\eta|$ & $r=16\epsilon$ & $\phi_e~$ & $n_s$& $N$ \\\hline
  30 &.00270 &.0617 & 1.0339 & .9715 & 38 \\\hline
  31 &.00253 &.0578 & 1.0328 & .9733 & 40 \\\hline
  32 &.00237& .0542 & 1.0317 & .9749 & 43\\\hline
  33 &.00223 &.0509 & 1.0308 & .9764 & 46\\\hline
  34 &.00210 &.0480 & 1.0299 & .9778 & 48 \\\hline
  35 &.00198 &.0453 & 1.0290 & .9790 & 51 \\\hline
  36 &.00187& .0428 & 1.0282 & .9802 & 54 \\\hline
  37 &.00177& .0405 & 1.0273 & .9812 & 57 \\\hline
  38 &.00168& .0384 & 1.0267 & .9822 &61 \\\hline
  \hline
\end{tabular}
\captionof{table}{$f(\phi) = \phi^{3\over 2}$, (case-2): ${\phi_b}=1.7$,\\ ${V_0}=-0.9\times {10^{-13}\mathrm{T}^{-2}},~{V_1}=0.9\times {10^{-13}\mathrm{T}^{-2}}.$}
      \label{table:7}
   \end{minipage}%
\hfill%
\begin{minipage}[h]{0.47\textwidth}
      \centering
\begin{tabular}{|c|c|c|c|c|c|}
  \hline\hline
  $\omega_0~$&$|\eta|$ & $r=16\epsilon$ & $\phi_e~$ & $n_s$& $N$ \\\hline
  29 &.00274 &.0594 & 1.0122& .9723 & 39 \\\hline
  30 &.00256 &.0555 & 1.0110& .9741 & 41 \\\hline
  31 &.00240&.0520 & 1.0097& .9757& 44\\\hline
  32 &.00225 &.0488 & 1.0090& .9772 & 47 \\\hline
  33 &.00212 &.0459 & 1.0080& .9786 & 50 \\\hline
  34 &.00199 &.0432 & 1.0071& .9798 & 53 \\\hline
  35 &.00188 &.0408 & 1.0063& .9809 & 56\\\hline
  36 &.00178 &.0386 & 1.0055& .9820 & 60 \\\hline
  37 &.00168& .0365 & 1.0048 & .9829 & 63\\\hline
  \hline
\end{tabular}
\captionof{table}{$f(\phi) = \phi^{3\over 2}$, (case-2): ${\phi_b}=1.7$,\\${V_0}=-0.9\times {10^{-13}\mathrm{T}^{-2}},~{V_1}=.92\times {10^{-13}\mathrm{T}^{-2}}.$}
      \label{table:8}
   \end{minipage}%
\end{figure}
\begin{figure}
\begin{minipage}[h]{0.47\textwidth}
\centering
\includegraphics[ width=0.9\textwidth] {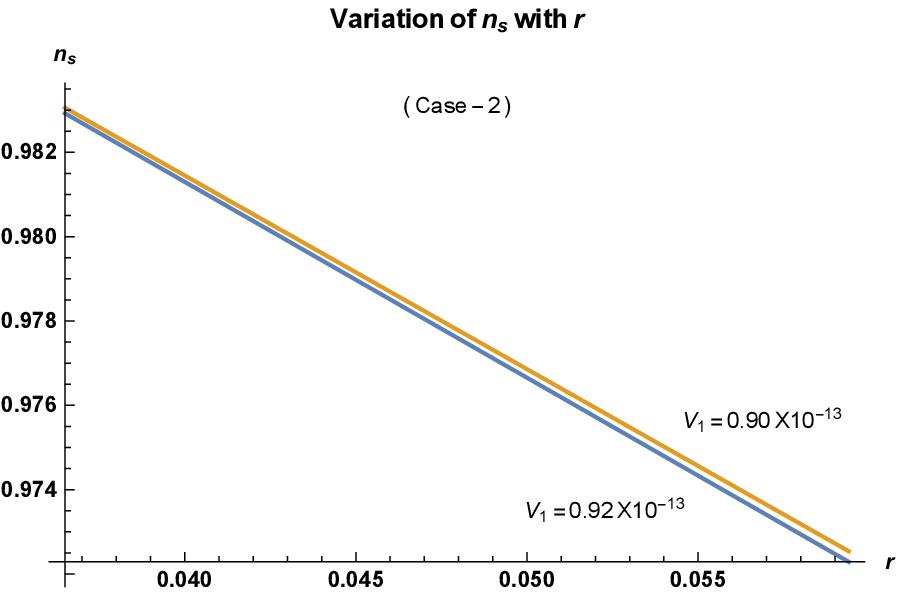}
 \caption{$f(\phi) = \phi^{3\over 2}$. Yellow ochre and blue colours represent table 7 and 8 respectively.}
      \label{fig:7}
   \end{minipage}%
\hfill%
\begin{minipage}[h]{0.47\textwidth}
\centering
\includegraphics[ width=0.9\textwidth] {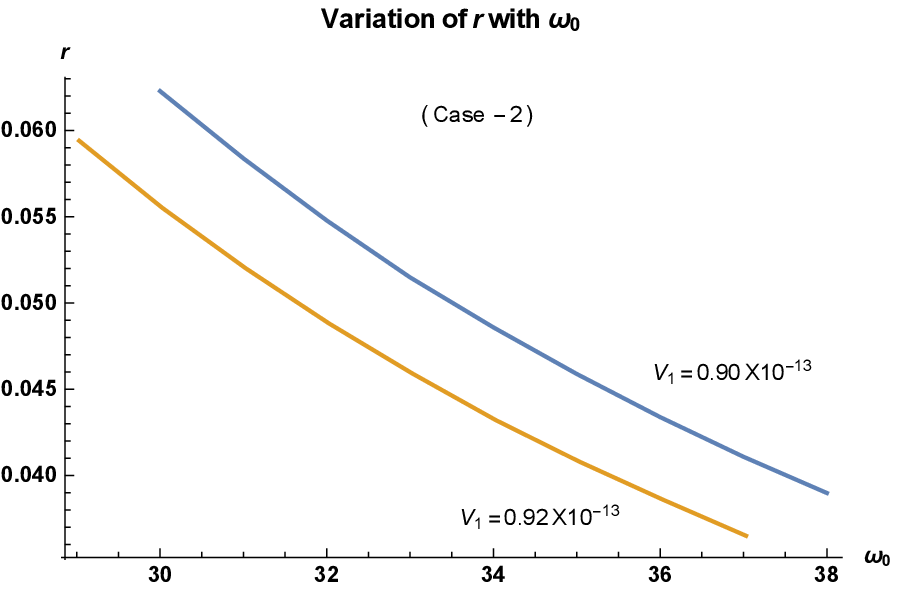}
 \caption{$f(\phi) = \phi^{3\over 2}$. Blue and yellow ochre colours represent table 7 and 8 respectively.}
      \label{fig:8}
   \end{minipage}%
\end{figure}
In order to study the behaviour of the scalar field at the end of inflation, we start with the Einstein's frame potential as before, $V_E=(V_1+{V_0\over\phi})$, and express the field equation (\ref{FE}) as,

\be\label{n11}\begin{split}& {3{H}^2\over V_1} = \frac{\dot\sigma^2}{2V_1} +\left(1+{V_0\over{ V_1\phi}}\right)\end{split}\ee
As the Hubble rate falls, and ${H}\ll V_1$, the above equation may be approximated to, ${\dot\sigma^2}=2i^2(V_1+{V_0\over\phi})$, which in terms of the scalar field $\phi$ reads as,

\be \dot \phi^2 + {2\over \omega_0^2}\big(V_0 + V_1 \phi\big)\phi^2 = 0,\ee
which may be solved to find

\be \phi = \frac{V_0}{V_1}\left[1-\tanh ^2\left\{-\frac{\sqrt{V_0}}{2} \left(c_1 + \frac{\sqrt{2}}{\omega_0}t\right)\right\}\right],\ee
which unfortunately is not oscillatory. Perhaps, due to the asymmetry of the potential, the oscillatory behaviour of the scalar field with an additive quadratic term is not exhibited.\\

\subsubsection{$n=2,~f(\phi) = \phi^2$}

\noindent
\textbf{Case-1:}
Under the choice $n = 2$, $f(\phi) = \phi^2$, and the potential \eqref{V} is now $V(\phi) = V_0 + V_1\phi^{4}$. Therefore the Brans-Dicke parameter, the Einstein's frame potential together with the slow-roll parameters \eqref{B-D} take the following forms,

\be \label{Para2}\begin{split}& \omega(\phi) = \frac{\omega_0^2 - 12\phi^2}{2\phi},~~~{d\sigma\over d\phi}={\omega_0\over \phi^2};\hspace{0.3 in} V_E = V_1 + V_0\phi^{-4},\hspace{0.3 in}\epsilon ={16 V_0^2\phi^2\over \omega_0^2(V_0 + V_1\phi^4)^2},\\&\eta = {24 V_0\phi^2\over \omega_0^2(V_0 + V_1\phi^4)},\hspace{0.3 in}N = \frac{\omega_0^2}{8\sqrt 2V_0}\left[{V_0\over 2}\left({1\over \phi_e^2} - {1\over \phi_b^2}\right) + {V_1\over 2}(\phi_b^2 - \phi_e^2) \right].\end{split}\ee
It is quite transparent that for large value of the scalar field $\phi$, a flat Einstein's frame potential is realizable here too. As before, we take two sets of data corresponding to two different values of $V_1 > 0$, and tabulate the parametric values in tables 9 and 10. One can see that the scalar tensor ratio $r \le 0.0202$, and the spectral index lies between $0.9645 \le n_s \le 0.9809$, which are in excellent agreement with Planck's data \cite{Planck181, Planck182}. Further number of e-folding $N$ is also sufficient to alleviate the flatness and the horizon problems. The $n_s$ versus $r$ and $r$ versus $\omega_0$ plots are presented in figures 9 and 10, as well.
\begin{figure}[h!]
\begin{minipage}[h]{0.47\textwidth}
      \centering
\begin{tabular}{|c|c|c|c|c|c|}
 \hline\hline
  $\omega_0~$&$|\eta|$ & $r=16\epsilon$ & $\phi_e~ $ & $n_s$& $N$ \\\hline
  26 &.013956 &.0202 & 1.0377 & .9645 & 37 \\\hline
  27 &.012941 &.0188 & 1.0364 & .9671 & 39 \\\hline
  28 &.012033 &.0175 & 1.0351& .9694 & 42\\\hline
  29 &.01122 &.0163 & 1.0339 & .9715 & 45 \\\hline
  30 &.010482 &.0152 & 1.0328 & .9733 & 49\\\hline
  31 &.009817 &.0142 & 1.0317 & .9750 & 52 \\\hline
  32 &.009213 &.0134 & 1.0308 & .9766 & 56 \\\hline
  33 &.008663 &.0126 & 1.0298 & .9780 & 59\\\hline
  34 &.008161 &.0118 & 1.0289 & .9792 & 63\\\hline
  35 &.007701 &.0112 & 1.0282 & .9804 & 67 \\\hline
   \hline
\end{tabular}
\captionof{table}{$f(\phi) = \phi^{2}$, (case-1): ${\phi_b}=1.7$,\\${V_0}=-0.9\times {10^{-13}\mathrm{T}^{-2}},~{V_1}=0.9\times {10^{-13}\mathrm{T}^{-2}}. $}
      \label{table:9}
   \end{minipage}%
\hfill%
\begin{minipage}[h]{0.47\textwidth}
      \centering
\begin{tabular}{|c|c|c|c|c|c|}
  \hline\hline
  $\omega_0~$&$|\eta|$ & $r=16\epsilon$ & $\phi_e~$ & $n_s$& $N$ \\\hline
  24 &.014543 &.0187 & 1.0127 & .9639 & 37\\\hline
  25 &.013403 &.0173 & 1.0112 & .9667 & 40 \\\hline
  26 &.012392 &.0160 & 1.0098 & .9692 & 43\\\hline
  27 &.011491 &.0148 & 1.0085 & .9714 & 46 \\\hline
  28 &.010685 &.0138 & 1.0073 & .9735 & 50\\\hline
  29 &.00996 &.0128 & 1.0062 & .9753& 54 \\\hline
  30 &.00931 &.0120 & 1.0051 & .9769 & 57\\\hline
  31 &.008717& .0112 & 1.0041 & .9784 & 61 \\\hline
  32 &.008180& .0105 & 1.0032 & .9797& 65 \\\hline
  33 &.007692& .0099 & 1.0023 & .9809 & 69 \\\hline
   \hline
\end{tabular}
\captionof{table}{$f(\phi) = \phi^{2}$, (case-1): ${\phi_b}=1.7$.\\${V_0}=-0.9\times {10^{-13}\mathrm{T}^{-2}},~{V_1}=1.0\times {10^{-13}\mathrm{T}^{-2}}.$}
      \label{table:10}
   \end{minipage}%
\end{figure}
\begin{figure}
\begin{minipage}[h]{0.47\textwidth}
\centering
\includegraphics[ width=0.9\textwidth] {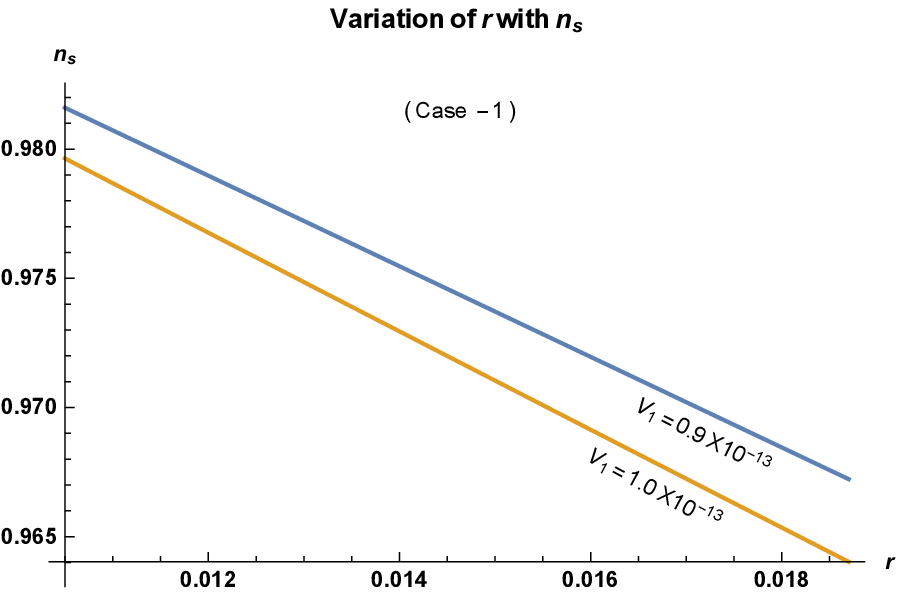}
 \caption{$f(\phi) = \phi^{2}$. Blue and yellow ochre colours represent table 9 and 10 respectively.}
      \label{fig:9}
   \end{minipage}%
\hfill%
\begin{minipage}[h]{0.47\textwidth}
\centering
\includegraphics[width=0.9\textwidth] {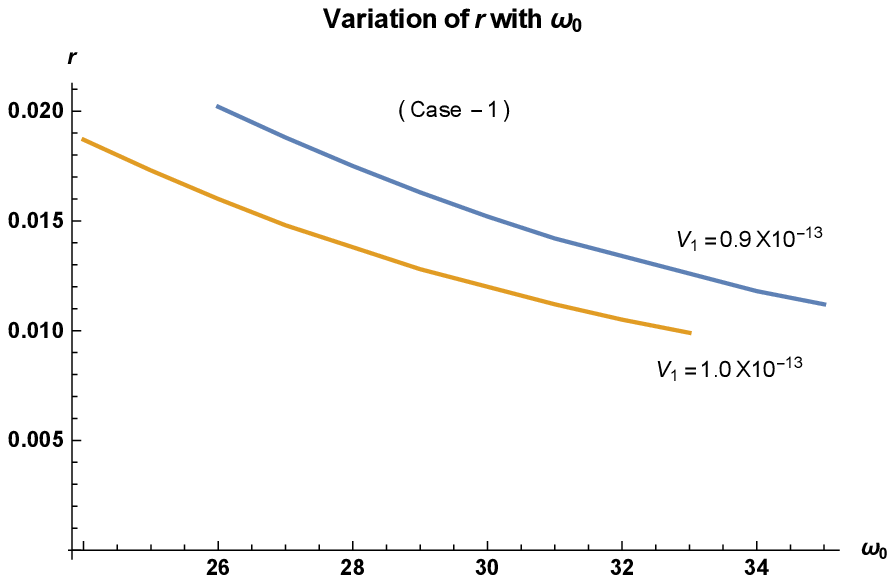}
 \caption{$f(\phi) = \phi^{2}$. Blue and yellow ochre colours represent table 9 and 10 respectively.}
      \label{fig:10}
   \end{minipage}%
\end{figure}
Equation \eqref{FE} now reads as,

\be 3H^2 = {1\over 2} \dot\sigma^2 + V_1 + {V_0\over \phi^4},\ee
which, as ${H}$ falls below $V_1$, i.e. ${H} \ll V_1$, may be approximated to, ${1\over 2}\dot\sigma^2 + {V_0\over\phi^4} + V_1 = 0$. In terms of the scalar field $\phi$ it is expressed as ${\dot\phi}^2 + {2 \over \omega_0^2}\left(V_1 \phi^4  + V_0\right) = 0$, since, $\dot\sigma = \omega_0{\dot\phi\over \phi^2}$, in view of \eqref{Para2}. The solution is,

\be \phi = - \frac{(-1)^{3/4} \sqrt[4]{V_0} \text{JacobiSN}\left(\left.c_1\sqrt[4]{-V_0 V_1} \left(1 + i {\sqrt{2}\over \omega_0}t\right)\right|-1\right)}{\sqrt[4]{V_1}},\ee
where JacobiSN is a meromorphic function in both arguments, which certain special arguments may automatically be evaluated to exact values.
In any case, under numerical simulation the above solution is found to exhibit oscillatory behaviour of the scalar field $\phi$. It is also clear that $\sigma = - {\omega_0\over \phi}$ oscillates as well, and the universe transits from inflationary regime to the matter dominated era.\\

\noindent
\textbf{Case-2:} Here, for $f(\phi)=\phi^2$, we consider the potential in the form, $V=V_1\phi^4+V_0\phi^2$, i.e. instead of a constant additive term, we consider $V_0\phi^2$ in addition. This case was earlier studied in \cite{Beh}. However, as already mentioned, in the years, Planck's data puts up tighter constraints on inflationary parameters, and so it is quite reasonable to check if this form of potential passes the said constraints \cite{Planck181, Planck182}. One can now find the expression for the Brans-Dicke parameter $\omega(\phi)$, the potential $V_E$ in the Einstein's frame, ${d\sigma\over d\phi}$, the slow-roll parameters $\epsilon, ~\eta$ and the number of e-folding $N$, in view of the equations \eqref{B-D1}, respectively as,

\be \label{ParaB}\begin{split}& \omega(\phi) = \frac{\omega_0^2 - 12\phi^2}{2\phi},~~~{d\sigma\over d\phi}={\omega_0\over \phi^2};\hspace{0.3 in} V_E = V_1+V_0\phi^{-2},\hspace{0.3 in}\epsilon ={4 V_0^2\phi^2\over \omega_0^2(V_1\phi^2+V_0)^2},\\&
\eta = {4V_0\phi^2\over \omega_0^2(V_1\phi^2+V_0)},\hspace{0.3 in}
N = \frac{\omega_0^2}{4\sqrt 2V_0}\left[{V_1}\ln(\phi_b - \phi_e)-{V_0\over 2}\left({1\over \phi_b^2} - {1\over \phi_e^2}\right)\right].\end{split}\ee
Note that the Einstein's frame potential now takes the same form as in case-1 for $n = 1$, and a flat section of the potential is still realizable at large value of the scalar field $\phi$. We present two tables 11 and 12, as before for different values of $V_1 >0 $. The scalar to tensor ratio $r \le 0.0636$ and the spectral index $0.9715 \le n_s \le 0.9831$ lie very much within the Planck's data, while the number of e-folding $N$ is again sufficient to alleviate the horizon and flatness problems. The plots (figures 11 and 12) represent $n_s$ versus $r$ and $r$ versus $\omega_0$ respectively. In view of the plots, the data for table 12, here appears to be even better.\\

\begin{figure}[h!]
\begin{minipage}[h]{0.47\textwidth}
      \centering
\begin{tabular}{|c|c|c|c|c|c|}
 \hline\hline
 ${\omega_0}$ &$|\eta|$& $r=16\epsilon$ & ${\phi_e}$ & $n_s$& $N$ \\\hline
 68 &.002337 &.0636 & 1.0148 & .9715 & 37 \\\hline
 70&.002206 &.0601 &1.0144 & .9731& 40 \\\hline
 72&.002085&.0568 &1.0140 & .9745& 42\\\hline
 74&.001974 &.0537 &1.0136 & .9759 & 44 \\\hline
 76&.001871 &.0509 & 1.0132 &.9772& 47 \\\hline
 78&.001776 &.0484 &1.0129 & .9783& 49 \\\hline
 80&.001689 &.0460 & 1.0126 &.9793 & 52 \\\hline
 82&.001607 &.0438 &1.0122 &.9804& 55 \\\hline
 84&.001532&.0417 & 1.0119 &.9813& 57\\\hline
 86&.001461&.0398 & 1.0117 &.9821& 60 \\\hline
 88&.001396 &.0380 & 1.0114 &.9830& 63 \\\hline
\hline
\end{tabular}
\captionof{table}{$f(\phi) = \phi^{2}$, (case-2): ${\phi_b}=1.26$.\\${V_0}=-1.0\times {10^{-20}\mathrm{T}^{-2}};~{V_1}=1.0\times {10^{-20}\mathrm{T}^{-2}}$.}
      \label{table:11}
   \end{minipage}%
\hfill%
\begin{minipage}[h]{0.47\textwidth}
      \centering
      \begin{tabular}{|c|c|c|c|c|c|}
      \hline\hline
       ${\omega_0}$&$|\eta|$ & $r=16\epsilon$ & ${\phi_e}$ & $n_s$& $N$ \\\hline
       60&.002363 &.0507 & 1.0656 & .9762& 38 \\\hline
       61&.002287 &.0490 & 1.0653 & .9770 & 39 \\\hline
       62&.002213 &.0475 & 1.0650 & .9778 & 40\\\hline
       63&.002144&.0460 & 1.0648 & .9785 & 41 \\\hline
       64 &.002078 &.0445 & 1.0646 & .9792& 43 \\\hline
       65 &.002014&.0432 & 1.0643 & .9798& 44 \\\hline
       66 &.001953&.0419 & 1.0641& .9804 & 46 \\\hline
       67&.001895&.0406 & 1.0638 & .9810 & 47 \\\hline
       68&.001840 &.0394 & 1.0636 & .9815 & 49 \\\hline
       69&.001787&.0383 & 1.0634 &.9821 & 50 \\\hline
       70&.001737&.0372 & 1.0632 &.9826 & 51 \\\hline
       71&.001688&.0362 &1.0630 &.9831 & 53 \\\hline
        \hline
\end{tabular}
\captionof{table}{$f(\phi) = \phi^{2}$, (case-2.): ${\phi_b}=1.26.$\\${V_0}=-1.0\times {10^{-20}\mathrm{T}^{-2}};{V_1}=1.1\times {10^{-20}\mathrm{T}^{-2}}$.}
      \label{table:12}
   \end{minipage}%
\end{figure}

\begin{figure}
\begin{minipage}[h]{0.47\textwidth}
\centering
 \includegraphics[width=0.9\textwidth] {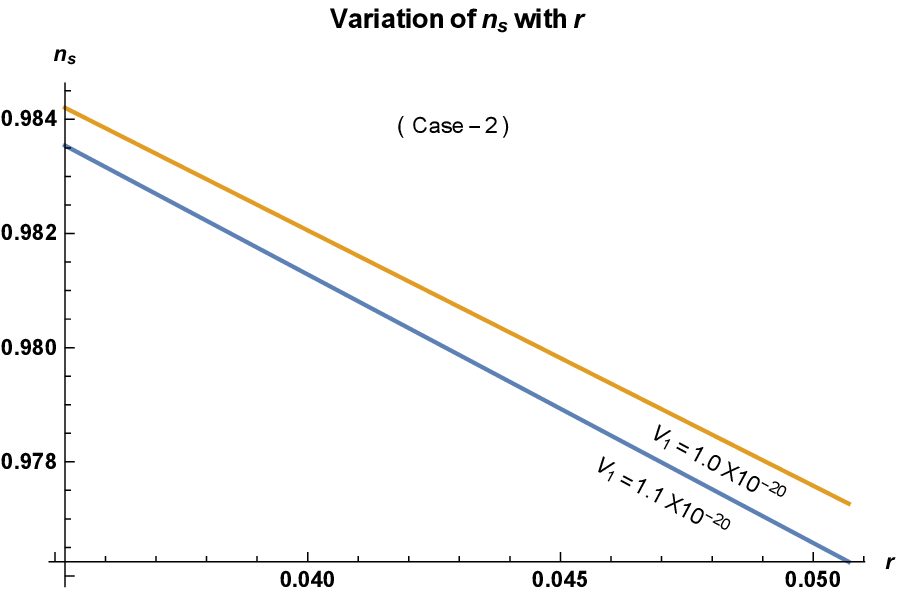}
 \caption{$f(\phi) = \phi^{2}$. Yellow ochre and blue  colours represent table 11 and 12 respectively.}
 \label{fig:11}
   \end{minipage}%
\hfill%
\begin{minipage}[h]{0.47\textwidth}
\centering
\includegraphics[width=0.9\textwidth] {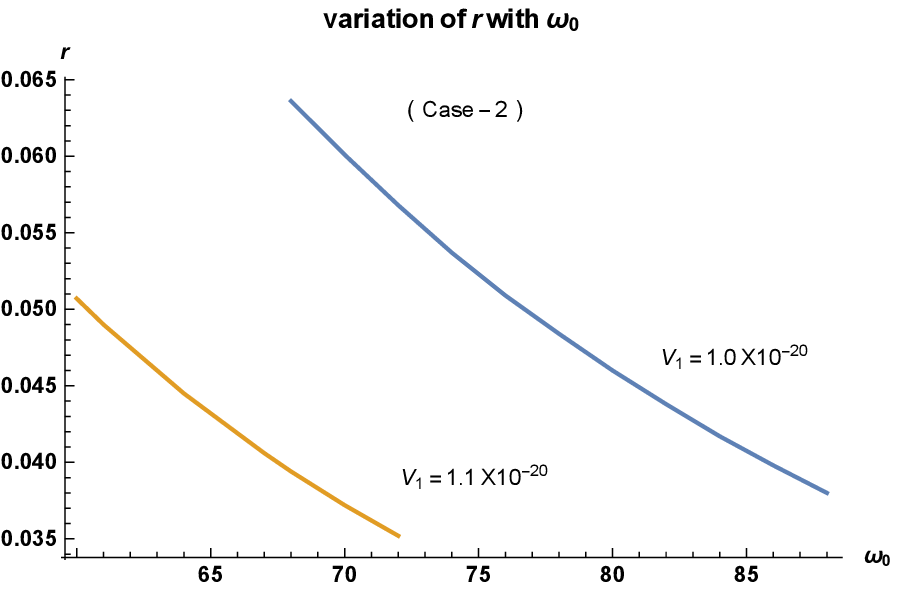}
 \caption{$f(\phi) = \phi^{2}$. Blue and yellow ochre colours represent table 11 and 12 respectively.}
      \label{fig:12}
   \end{minipage}%
\end{figure}

\noindent
To check if the scalar field executes oscillatory behaviour at the end of inflation, we note that here $V_E = V_1+{V_0\over\phi^2}$. So in view of equation (\ref{FE}) one obtains,

\be\label{n20}\begin{split}& {3{H}^2\over V_1} = \frac{\dot\sigma^2}{2V_1} +\left(1 +{ V_0\over V_1\phi^2}\right)\end{split}\ee
As, ${H}$ falls below $V_1$, and ${H} \ll V_1$, the above equation can be approximated to, ${\dot\sigma^2}=2i^2\big(V_1+{V_0\over\phi^2}\big)$, yielding, ${\dot\phi}=i{\phi\over\omega_0}\sqrt{2(V_0+{V_1\phi^2}})$, where, $\dot\sigma = \omega_0{\dot\phi\over \phi^2}$. Thus, we obtain,

\be{\phi\over{\left(\sqrt V_0+\sqrt{(V_0+V_1\phi^2)}\right)}}= {\sqrt V_0}e^{i{\sqrt {2V_0}\over\omega_0}t}.\ee
It is also possible to solve for $\phi$ and express it in the following form,

\be\begin{split} &\phi = \frac{\sqrt{-b_0 + \sqrt{b_0^2 - 4 a_0 c_0}}}{\sqrt{2a_0}},\\&
\mathrm{where},~~
a_0 = 1 - V_0V_1 e^{i 2{\sqrt{2V_0}\over \omega_0}t}, \hspace{0.2 cm} b_0 = -4V_0^3V_1 e^{i 4{\sqrt{2V_0}\over \omega_0}t}, \hspace{0.2 cm} c_0 = -4V_0^4 e^{i 4{\sqrt{2V_0}\over \omega_0}t}.\end{split}\ee
It is now quite apparent that $\phi$ executes oscillatory behaviour and therefore graceful exit from inflation is realizable. Since in view of \eqref{ParaB} $\sigma = - {\omega_0\over \phi}$, therefore $\sigma$ also executes oscillatory behaviour.\\

\noindent
\textbf{Case-3:} We consider yet another case for $n = 2$, i.e. taking $f(\phi)=\phi^2$, with the potential $V(\phi)$ being represented by two additional terms apart from $\phi^4$, as it should be, to make it a perfect square: $V(\phi) = (\sqrt{V_1}\phi^2 - \sqrt{V_0}\phi)^2 = V_1\phi^4 + V_0\phi^2 - 2\sqrt{V_0V_1} \phi^3$. As before, one can now find the expression for the Brans-Dicke parameter $\omega(\phi)$, the potential $V_E$ in the Einstein's frame, ${d\sigma\over d\phi}$, the slow-roll parameters $\epsilon, ~\eta$ and the number of e-foldings $N$, in view of the equations \eqref{2.9}, \eqref{3.3} and \eqref{3.4} respectively as,

\be \label{ParaS}\begin{split}& \omega(\phi) = \frac{\omega_0^2 - 12\phi^2}{2\phi},~~~{d\sigma\over d\phi}={\omega_0\over \phi^2};\hspace{0.3 in} V_E = V_1-2{\sqrt{V_0 V_1}\over\phi}+{V_0\over\phi^2},\hspace{0.3 in}\epsilon ={4 V_0^2 \phi^2 \over \omega_0^2(\sqrt{V_0 V_1}\phi-V_0)^2},\\&
\eta ={4 V_0^2 \phi^2 \over \omega_0^2(\sqrt{V_0 V_1}\phi-V_0)^2},\hspace{0.3 in}
N =\int_{\phi_e}^{\phi_b}{\omega_0^2\over 4\sqrt 2V_0}\frac{(\sqrt{V_0 V_1}\phi-V_0)}{\phi^3} d\phi .\end{split}\ee
One can clearly see that the flat section of the potential is still attainable for large value of the scalar field $\phi$. Table 13 and Table 14 depict that the scalar to tensor ratio $r < 0.1$), is quite reasonable, while the spectral index $0.9752 \le n_s \le 0.981$ fits perfectly with Planck's data \cite{Planck181, Planck182}. Figures 13, and 14 represent $n_s$ versus $r$ and $r$ versus $\omega_0$ plots respectively. Interestingly, two $r$ versus $n_s$ plots (figure-13) corresponding to the two sets of data (Table-13 and Table-14) merge almost perfectly.\\

\begin{figure}[h!]
\begin{minipage}[h]{0.47\textwidth}
      \centering
      \begin{tabular}{|c|c|c|c|c|c|} 
       \hline\hline
       ${\omega_0} $&$\eta$ & $r=16\epsilon$ & ${\phi_e}$ & $n_s$& $N$ \\\hline
       110 &.006208&.0993 &1.0185 & .9752 & 57 \\\hline
       112 &.005988&.0958  &1.0182 & .9760& 59 \\\hline
       114 &.005780& .0925 &1.0178 & .9769& 61 \\\hline
       116 &.005582 &.0893 &1.0175  & .9777& 63 \\\hline
       118 &.005394&.0863 & 1.0172 & .9784& 65 \\\hline
       120 &.005216&.0835 &1.0170 & .9791& 67 \\\hline
       122 &.005046&.0807 & 1.0167& .9798& 70 \\\hline
       124 &.004885& .0782& 1.0164 & .9804 & 72 \\\hline
       126 &.004731& .0757&1.0161 & .9810 & 74 \\\hline
       \hline
\end{tabular}
\captionof{table}{$f(\phi) = \phi^{2}$, (case-3): ${\phi_b}=1.3$.\\${V_1}=0.9\times {10^{-20}\mathrm{T}^{-2}};~{V_0}=0.9\times {10^{-20}\mathrm{T}^{-2}}$.}
      \label{tab:13}
   \end{minipage}%
   \hfill%
\begin{minipage}[h]{0.47\textwidth}
      \centering
      \begin{tabular}{|c|c|c|c|c|c|} 
      \hline\hline
      ${\omega_0}$&$\eta$ & $r=16\epsilon$ & ${\phi_e}$ & $n_s$& $N$ \\\hline
       142&.006160 &.0986 &1.0386 & .9754 & 57 \\\hline
       144&.005990 & .0958 &1.0388 & .9760& 59 \\\hline
       146&.005827& .0932 &1.0391 & .9767& 60\\\hline
       148&.005671& .0907 &1.0393  & .9773& 62 \\\hline
       150&.005521&.0883 & 1.0395 & .9779& 64 \\\hline
       152 &.005376&.0860 &1.0397 & .9785& 65 \\\hline
       154 &.005237&.0838 & 1.0399& .9791 & 67 \\\hline
       156&.005104 &.0817& 1.0400& .9796 & 69\\\hline
       158 &.004976&.0796& 1.0402 & .9801 & 71 \\\hline
       160 &.004852&.0776& 1.0404 & .9806 & 72 \\\hline
       \hline
\end{tabular}
\captionof{table}{$f(\phi) = \phi^{2}$, (case-3): ${\phi_b}=1.3$.\\${V_1}=1.0\times {10^{-20}\mathrm{T}^{-2}};~{V_0}=0.9\times {10^{-20}\mathrm{T}^{-2}}$.}
      \label{tab:14}
   \end{minipage}%
\end{figure}
\begin{figure}
\begin{minipage}[h]{0.47\textwidth}
\centering
\includegraphics[width=0.9\textwidth] {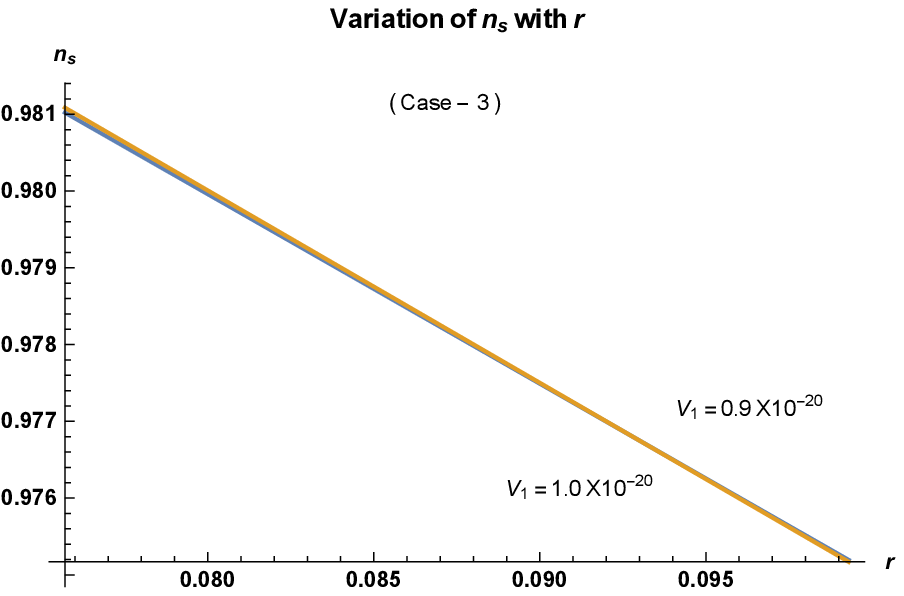}
 \caption{$f(\phi) = \phi^{2}$. Blue and yellow ochre colours represent table 13 and 14 respectively.}
      \label{fig:13}
   \end{minipage}%
 \hfill%
\begin{minipage}[h]{0.47\textwidth}
\centering
\includegraphics[width=0.9\textwidth] {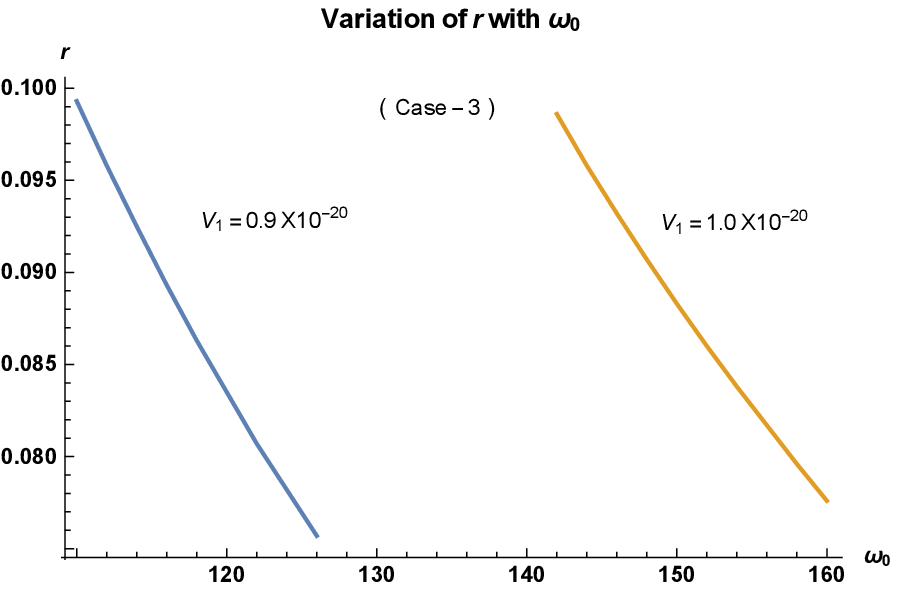}
 \caption{$f(\phi) = \phi^{2}$. Blue and yellow ochre colours represent table 13 and 14 respectively.}
      \label{fig:14}
   \end{minipage}%
\end{figure}

\noindent
The scalar field executes oscillatory behaviour here to, as we demonstrate below. Here, $V_E = V_1-2{\sqrt{V_0 V_1}\over\phi}+{V_0\over\phi^2}$, and so from (\ref{FE}) we find,

\be\label{n21}\begin{split}& {3{H}^2\over V_1} = \frac{\dot\sigma^2}{2V_1}+\left(1 -{2\over \phi}{\sqrt{V_0\over V_1}}\right) + {V_0\over V_1\phi^2}. \end{split}\ee
As ${H}$ falls below $V_1$, and  ${H} \ll V_1 $, the above equation can be approximated as,
${\dot\sigma^2}=2i^2{({\sqrt{ V_0}\over \phi}-{\sqrt V_1})^2}$, which yields ${\dot\phi}=i{\phi\over\omega_0}{\sqrt 2({\sqrt V_0}-{\sqrt V_1}\phi)}$ where,$\dot\sigma = \omega_0{\dot\phi\over \phi^2}$.
Therefore finally we obtain,

\be{\phi} = \frac{{\sqrt V_0}e^{i{\sqrt{2V_0}\over \omega_0}t}}{1 + \sqrt{V_1}e^{i{\sqrt{2V_0}\over \omega_0}t}} .\ee
Clearly, $\phi$ executes oscillatory behaviour, and graceful exit from inflation may be realized hereto. Here again since in view of \eqref{ParaS} $\sigma = - {\omega_0\over \phi}$, therefore $\sigma$ executes oscillatory behaviour, as well.

\subsection{Exponential potential:} Finally, we consider an exponential form of the potential with: $f(\phi) = e^{\lambda\phi\over 2}$, with $V(\phi) = V_1e^{\lambda\phi}+ V_0$. It is possible to find the expression for the Brans-Dicke parameter $\omega(\phi)$, the potential $V_E$ in the Einstein's frame, ${d\sigma\over d\phi}$, the slow-roll parameters $\epsilon, ~\eta$ and the number of e-folding $N$, in view of the equations \eqref{2.9}, \eqref{3.3} and \eqref{3.4} respectively as,

\be \label{ParaE}\begin{split}& \omega(\phi) = \frac{(\omega_0^2 -{3\over 4}\lambda^2 e^{\lambda\phi})\phi}{2e^{\lambda\phi\over 2}};~~~{d\sigma\over d\phi}={\omega_0\over e^{\lambda\phi\over 2}};\hspace{0.3 in} V_E=V_1+V_0e^{-\lambda\phi},\hspace{0.3 in}\epsilon = {\lambda^2 V_0^2 e^{\lambda\phi}\over \omega_0^2(V_1e^{\lambda\phi}+V_0)^2},\\&\eta = {\lambda^2 V_0 e^{\lambda\phi}\over \omega_0^2(V_1e^{\lambda\phi}+V_0)}, \hspace{0.3 in}
N =\int_{\phi_e}^{\phi_b}\frac{\omega_0^2(V_0+V_1e^{\lambda\phi})}{2{\sqrt 2} V_0\lambda e^{\lambda\phi}} d\phi.\end{split}\ee
We present our results in the following table 15, under the only choice of the parameter $\lambda = -1$. The data shows good agreement $r < 0.06$, and $0.9612 \le n_s \le 0.9836$ with Planck's data \cite{Planck181, Planck182}. Figure 15, represents a plot for $r$ versus $n_s$.
\begin{figure}[h!]
\begin{minipage}[h]{0.47\textwidth}
      \centering
\begin{tabular}{|c|c|c|c|c|c|}
  \hline\hline
 ${\omega_0 }$ &$\eta$ & $r=16\epsilon$ & ${|\phi_e|} $ & $n_s$& $N$ \\\hline
13 &.008468& .0584 & .07690 &.9612 &30 \\\hline
14 &.007301&.0503 &.07141 & .9665 &34 \\\hline
15 &.00636&.0439 &.06665 & .9708 &40 \\\hline
16 &.005589&.0385 & .06249 &.9745 & 45\\\hline
17&.004952&.0341&.05882& .9773 &51 \\\hline
18 &.004417& .0305 &.05555 &.9797 &57  \\\hline
19&.003964&.02734 &.05263 & .9818 &64 \\\hline
20 &.003578& .02467 & .04999 &.9836 &71 \\\hline
\hline
\end{tabular}
\captionof{table}{$|{\phi_b}|=1.2,{V_0}=-1.0\times {10^{-20}\mathrm{T}^{-2}},\\{V_1}=1.0\times {10^{-20}\mathrm{T}^{-2}}$.}
      \label{table:15}
   \end{minipage}%
\hfill%
\begin{minipage}[h]{0.47\textwidth}
\centering
\includegraphics[width=0.9\textwidth] {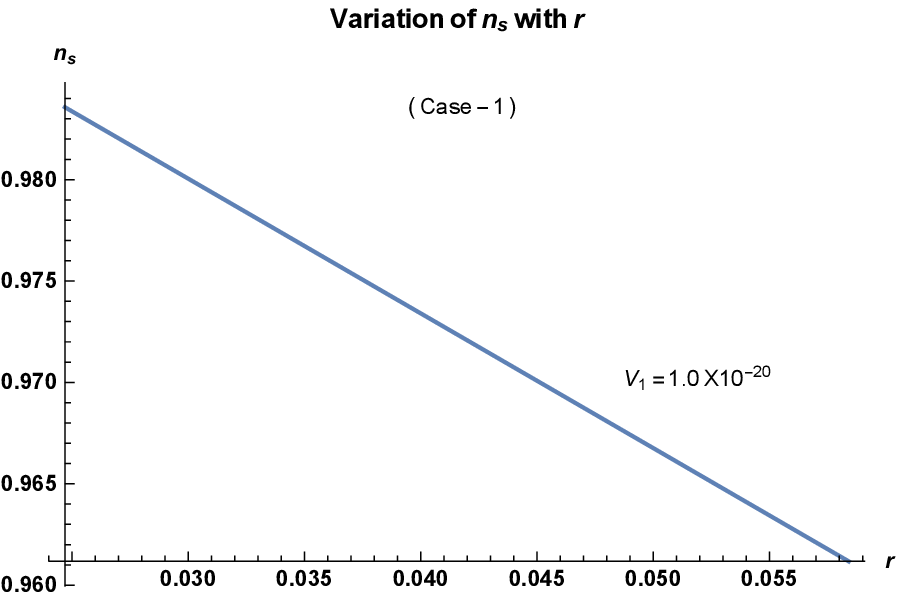}
 \caption{$f(\phi) = e^{-{\phi\over 2}}$}
      \label{fig:15}
   \end{minipage}%
\end{figure}
The scalar field executes oscillatory behaviour here to, as we demonstrate below. Here $V_E =V_1+V_0e^{\phi} $, and so in view of equation (\ref{FE}), one can calculate,

\be\label{exp}\begin{split}& {3{H}^2\over V_1} = \frac{\dot\sigma^2}{2V_1} +\left(1+{V_0\over V_1} e^{\phi}\right)\end{split}\ee
Again as ${H}$ falls below $V_1$, and  ${H} \ll V_1 $, the above equation can be approximated to,
${\dot\sigma^2}=2i^2(V_1+V_0e^{\phi})$, which yields ${\dot\phi}=i{e^{-{\phi\over 2}}\over\omega_0 }\sqrt{2(V_1+V_0e^{\phi})}$, where,$\dot\sigma = {\omega_0} \dot\phi e^{\phi\over 2}$. Finally, therefore

\be \phi = \ln{\left[{1\over 4V_0^2}\left(e^{i{\sqrt{2V_0}\over \omega_0}t} - 2V_0V_1 +V_0^2 V_1^2 e^{-i{\sqrt{2V_0}\over \omega_0}t} \right)\right]}.\ee
The oscillatory behaviour of the scalar field here again assures graceful exit from inflationary regime. In view of \eqref{ParaE}, $\sigma = 2\omega_0 e^{\phi\over 2}$, for $\lambda = -1$, which we have considered, hence $\sigma$ also executes oscillatory behaviour.

\section{\bf{Concluding remarks}}

Scalar-tensor theories of gravity are generalizations of the Brans-Dicke theory, in which the coupling parameter is a function of the scalar field, i.e. $\omega_{BD} = \omega(\phi)$, and therefore is a variable. The requirement for such generalization of Brans-Dicke theory generated from the tight constraints on $\omega_{BD}$ established by the solar system experiments \cite{C01}. There exists various classification of scalar-tensor theory of gravity \cite{C02}. In the present manuscript we have considered standard non-minimal coupling, where the coupling parameter $f(\phi)$ is arbitrary. It has been noticed earlier that such a theory has an in-built symmetry being associated with a conserved current for trace-free fields, such as vacuum and radiation dominated eras for barotropic fluids. In view of such a symmetry it is possible to fix all the variables of the theory, including the potential function, fixing the form of one of those. In this manuscript, we have chosen different forms of the coupling parameter $f(\phi)$, which fixed $\omega(\phi)$ and $V(\phi)$, to study the cosmological evolution of the very early universe in the context of inflation. Inflation is a quantum mechanical phenomenon, and has occurred around Planck's era. However, it has been argued that since the radiative corrections to the potential are negligible, hence the inflationary parameters can be computed using the classical Lagrangian \cite{C03}. This argument leads in general, to calculate inflationary parameters in view of the classical Lagrangian, which we have done in the present manuscript. The so called unification programmes, which essentially claim to unify early inflationary regime with late-time cosmic acceleration have no credentials, since none of the models passes through a well behaved radiation and early matter dominated era. However, a history of cosmic evolution starting from inflationary regime, followed by a Friedmann-like radiation ($a \propto \sqrt t$) and early matter dominated eras ($a \propto t^{2\over 3}$), that finally ends up to a late-time accelerated unverse ($z = 0.75$), has already been explored in view of the present \cite{Beh}. In this connection, the present model makes a reasonably viable attempt to unify early inflation with late-time cosmic acceleration. Nevertheless earlier, only a single form of the coupling parameter $f(\phi)$ together with a particular form of $V(\phi)$ had been treated. Here, we have extended our work considering at least three power potentials together with an exponential potential. We find that quadratic, cubic, quartic and exponential potential pass the tighter constraints on inflationary parameters released by latest Planck's data \cite{Planck181, Planck182} comfortably. Further, all these potentials admit graceful exit from inflation, except one case of cubic potential associated with a square potential, for which unfortunately the scalar field does not show up oscillatory behaviour at the end of inflation.\\

For the purpose of the present analysis, we have translated the non-minimally coupled Jordan's frame action to the Einstein's frame, under conformal transformation. It is therefore worth to make certain comments in this regard. There is an age old debate regarding physical equivalence between the two: Jordan's and Einstein's frames, which are related under conformal transformation. Now, indeed if the two formulations are not equivalent, the problem arises in selecting the physically preferred frame. It emerges from the work of several authors, in different contexts on Kaluza-Klein and Brans-Dicke theories, that the formulations of a scalar-tensor theory in the two conformal frames are physically inequivalent  \cite{C1, C2, C3, C4, C5, C6, C7, C7a}. Also the Jordan frame formulation of a scalar-tensor theory is not viable because the energy density of the gravitational scalar field present in the theory is not bounded from below, which amounts to the violation of the weak energy condition \cite{C8}). The system therefore is unstable and decays toward a lower and lower energy state ad infinitum \cite{C6, C7, C7a}. Although, a quantum system may have states with negative energy density \cite{C8, C9, C9a}, such feature is not acceptable for a viable classical theory of gravity. In fact, a classical theory must have a ground state that is stable against small perturbations. The violation of the weak energy condition by scalar-tensor theories formulated in the Jordan conformal frame makes them unviable descriptions of classical gravity, while the Einstein frame formulation of scalar-tensor theories is free from such problem. However, in the Einstein frame also there is a violation of the equivalence principle due to the anomalous coupling of the scalar field to ordinary matter. Nevertheless, this violation is small and compatible with the available tests of the equivalence principle \cite{C10}. Further, Einstein's frame is indeed regarded as an important low energy manifestation of compactified theories \cite{C10, C11, C12, C13, C14, C15}. However, in search of Noether symmetries of $F(R)$ theory of gravity, the two frames have been found to be physically equivalent \cite{C16}. So although the debate persists, but somehow it is quite relevant to consider Einstein's frame to be the physical frame. Therefore, In view of the above discussions, it is justified to study the physics associated with non-minimally coupled scalar-tensor theory of gravity, after translating it to the Einstein's frame, as we have done in the present article.

\end{document}